\documentclass[%
reprint, hidelinks,
superscriptaddress,
hypertext,
showkeys,
showpacs,
 nofootinbib,
 nobibnotes,
 amsmath,amssymb,
 aps,
prc,
floatfix,
]{revtex4-2}

\usepackage{graphicx}% Include figure files
\usepackage{dcolumn}% Align table columns on decimal point
\usepackage{bm}% bold math
\usepackage{isotope}
\usepackage{upgreek}
\usepackage{xcolor}
\usepackage{longtable}
\usepackage[caption=false]{subfig}
\usepackage{booktabs}
\usepackage{tablefootnote}
\usepackage{threeparttable}
\usepackage{epigraph}
\usepackage{comment}
\usepackage[]{hyperref}% add hypertext capabilities
\hypersetup{
  colorlinks   = true, %Colours links instead of ugly boxes
  urlcolor     = magenta, %Colour for external hyperlinks
  linkcolor    = magenta, %Colour of internal links
  citecolor   =  cyan %Colour of citations
}
\usepackage[mathlines]{lineno}% Enable numbering of text and display math
\newcommand{\beag}{$^{7}$Be$(\alpha,\gamma)^{11}$C}

\begin{document}

\preprint{APS/123-QED}

\title{ First inverse kinematics measurement of resonances
 in $^7$Be($\alpha,\gamma$)$^{11}$C relevant to
 neutrino--driven wind nucleosynthesis using DRAGON}% Force line breaks with \\

\author{A. Psaltis}
\email{psaltisa@mcmaster.ca}
\altaffiliation[Present address: ]{Institut f\"ur Kernphysik,
Technische Universit\"at Darmstadt, 64289 Darmstadt, Germany
}
\affiliation{
 Department of Physics \& Astronomy, McMaster University,
 Hamilton, Ontario L8S 4M1, Canada
}
\affiliation{
  The NuGrid collaboration \href{https://nugrid.github.io/}{https://nugrid.github.io/}
}
\author{A.A. Chen}
\affiliation{
 Department of Physics \& Astronomy, McMaster University,
 Hamilton, Ontario L8S 4M1, Canada
}
\affiliation{
 The NuGrid collaboration \href{https://nugrid.github.io/}{https://nugrid.github.io/}
}
\author{R. Longland}
\affiliation{Department of Physics, North Carolina State University,
Raleigh, NC, 27695, USA}
\affiliation{Triangle Universities Nuclear Laboratory, Duke University, Durham, NC, 27710, USA}
\author{D.S. Connolly}
\altaffiliation[Present address: ]{Los Alamos National Laboratory, \\
Los Alamos, NM 87545, USA}
\affiliation{TRIUMF, 4004 Wesbrook Mall, Vancouver,
British Columbia V6T 2A3, Canada}
\author{C.R. Brune}
\affiliation{Department of Physics \& Astronomy,
Ohio University, Athens, Ohio 45701, USA}
\author{B. Davids}
\affiliation{TRIUMF, 4004 Wesbrook Mall, Vancouver,
British Columbia V6T 2A3, Canada}
\affiliation{Department of Physics, Simon Fraser University, Burnaby, British Columbia V5A 1S6, Canada}
\author{J. Fallis}
\affiliation{North Island College, 2300 Ryan Rd, Courtenay,
BC V9N 8N6, Canada}
\author{R. Giri}
\affiliation{Department of Physics \& Astronomy,
Ohio University, Athens, Ohio 45701, USA}
\author{U. Greife}
\affiliation{Department of Physics, Colorado School of Mines,
Golden, Colorado 80401, USA}
\author{D.A. Hutcheon}
\affiliation{TRIUMF, 4004 Wesbrook Mall, Vancouver,
British Columbia V6T 2A3, Canada}
\author{L. Kroll}
\affiliation{
 Department of Physics \& Astronomy, McMaster University,
 Hamilton, Ontario L8S 4M1, Canada
}
\affiliation{
   The NuGrid collaboration \href{https://nugrid.github.io/}{https://nugrid.github.io/}
}
\author{A. Lennarz}
\affiliation{TRIUMF, 4004 Wesbrook Mall, Vancouver,
British Columbia V6T 2A3, Canada}
\author{J. Liang}
\altaffiliation[Present address: ]{TRIUMF, 4004 Wesbrook Mall, Vancouver,
British Columbia V6T 2A3, Canada
}
\affiliation{
 Department of Physics \& Astronomy, McMaster University,
 Hamilton, Ontario L8S 4M1, Canada
}
\author{M. Lovely}
\affiliation{Department of Physics, Colorado School of Mines,
Golden, Colorado 80401, USA}
\author{M. Luo}
\affiliation{Department of Physics \& Astronomy, University of British
Columbia, Vancouver, British Columbia V6T 1Z4, Canada}
\author{C. Marshall}
\altaffiliation[Present address: ]{Department of Physics \& Astronomy,
Ohio University, Athens, Ohio 45701, USA}
\affiliation{Department of Physics, North Carolina State University,
Raleigh, NC, 27695, USA}
\affiliation{Triangle Universities Nuclear Laboratory, Duke University, Durham, NC, 27710, USA}

\author{S.N. Paneru}
\altaffiliation{Facility for Rare Isotope Beams, East Lansing, Michigan 48824, USA}
\affiliation{Department of Physics \& Astronomy,
Ohio University, Athens, Ohio 45701, USA}
\author{A. Parikh}
\affiliation{Department de F\'isica, Universitat Polit\`ecnica de Catalunya, E-08036 Barcelona, Spain}
\author{C. Ruiz}
\affiliation{TRIUMF, 4004 Wesbrook Mall, Vancouver,
British Columbia V6T 2A3, Canada}
\affiliation{Department of Physics \& Astronomy, University of Victoria,
Victoria, BC  V8W 2Y2, Canada}
\author{A.C. Shotter}
\affiliation{School of Physics, University of Edinburgh EH9
3JZ Edinburgh, United Kingdom}
\author{M. Williams}
\affiliation{TRIUMF, 4004 Wesbrook Mall, Vancouver,
British Columbia V6T 2A3, Canada}
\affiliation{Department of Physics, University of York,
Heslington, York YO10 5DD, United Kingdom}

\begin{abstract}

A possible mechanism to explain the origin of the light \emph{p}--nuclei in the
Galaxy is the nucleosynthesis in the proton--rich neutrino--driven wind ejecta of
core--collapse supernovae via the $\nu p$--process. However this production
scenario is very sensitive to the underlying supernova dynamics and the nuclear
physics input. As far as the nuclear uncertainties are concerned, the breakout
from the \emph{pp}-chains via the
\beag~ reaction has been identified as an important link which can influence the nuclear flow
and therefore the efficiency of the $\nu p$--process. However
its reaction rate is poorly known over the relevant temperature range, T = 1.5--3~GK. We report on the first direct measurement of two resonances of
the \beag~reaction with previously
unknown strengths using an intense radioactive $\isotope[7][]{Be}$ beam from the
ISAC facility and the DRAGON recoil separator in inverse kinematics.
We have decreased the \beag~reaction rate uncertainty to $\sim 9.4-10.7$\% over the
relevant temperature region.
\end{abstract}

%\begin{description}
%\end{description}

\maketitle

%\tableofcontents

\section{Introduction}
\label{sec:1}

The origin of the roughly 35 neutron--deficient stable isotopes with masses
$A \geq 74$ --- between $\isotope[74][]{Se}$ and $\isotope[196][]{Hg}$ --- in the
proton--rich side of the valley of stability, known as the ``\textit{p}--nuclei'' is a
long--standing puzzle in nuclear astrophysics~\cite{arnould2003p, rauscher2013constraining,pignatari2016production}.
The \textit{p}--nuclei were also traditionally referred to as ``excluded'' nuclei, since they were
``shielded'' by the \emph{s}-- and the \emph{r}--process reaction paths~\cite{cameron1957nuclear}. For this reason
their observed solar abundances~\cite{lodders20094}, are 1--2 orders of magnitude smaller
than their \textit{s}-- and \textit{r}--process counterparts in the same mass region.
It is generally accepted that the \emph{p}--nuclei in the solar system
have been produced by more than one process; however their synthesis mechanism
is commonly referred to as ``\emph{p}--process''.

The photodisintegration of pre--existing neutron--rich seeds, which is one of the most
promising nucleosynthesis scenarios of \textit{p}--nuclei synthesis and is thought to take
place in the oxygen--neon layer of core--collapse supernovae (ccSNe),
cannot reproduce the solar abundances of the light $\isotope[92,94][]{Mo}$ and
$\isotope[96,98][]{Ru}$ isotopes, as well as the rare species $\isotope[113][]{In}$, $\isotope[115][]{Sn}$ and
$\isotope[138][]{La}$. Additional astrophysical sites/nucleosynthesis scenarios have been
proposed, such as the thermonuclear explosions of Chandrasekhar mass
carbon--oxygen white dwarfs (CO WD)~\cite{travaglio2014radiogenic}, which is
also supported by Galactic Chemical Evolution (GCE) models and the
\textit{rp}--process in Type I X--ray bursts~\citep{schatz1998rp}.
It is remarkable that despite the variety of astrophysical models, all
these processes can reproduce the solar abundances of most of the
\emph{p}--nuclei to within a factor of 3 \citep[\textit{e.g.} see the sensitivity
studies in References][]{rapp2006sensitivity, rauscher2016uncertainties}.

The advancement of multi--dimensional core--collapse supernova simulations
with sophisticated neutrino transport methods, see References~\cite{PhysRevLett.104.251101, 2010A&A...517A..80F} for the first studies that discussed this and~\cite{Just:2015, OConnor:2015} for some more recent results,
suggests that the composition of the early innermost ejecta of the
neutrino--driven wind that drives the explosion are mostly proton--rich (the electron fraction $Y_e$ is greater than 0.5)\footnote{A relatively neutron--rich neutrino--driven wind ($0.4 < Y_e < 0.5$), leads to a different nucleosynthesis scenario called the weak \textit{r}--process, which can produce the lighter heavy elements with Z = 26--47~\cite{arcones2011production}.}~\cite{wanajo2018nucleosynthesis, vartanyan2019successful},
and that gives rise to a new nucleosynthesis scenario, the
$\nu p$--process~\cite{frohlich2006neutrino,wanajo2006rp,pruet2006nucleosynthesis}, which can produce the lighter of the \textit{p}--nuclei.

\begin{figure}[ht!]
    \centering
    \includegraphics[width=.5\textwidth]{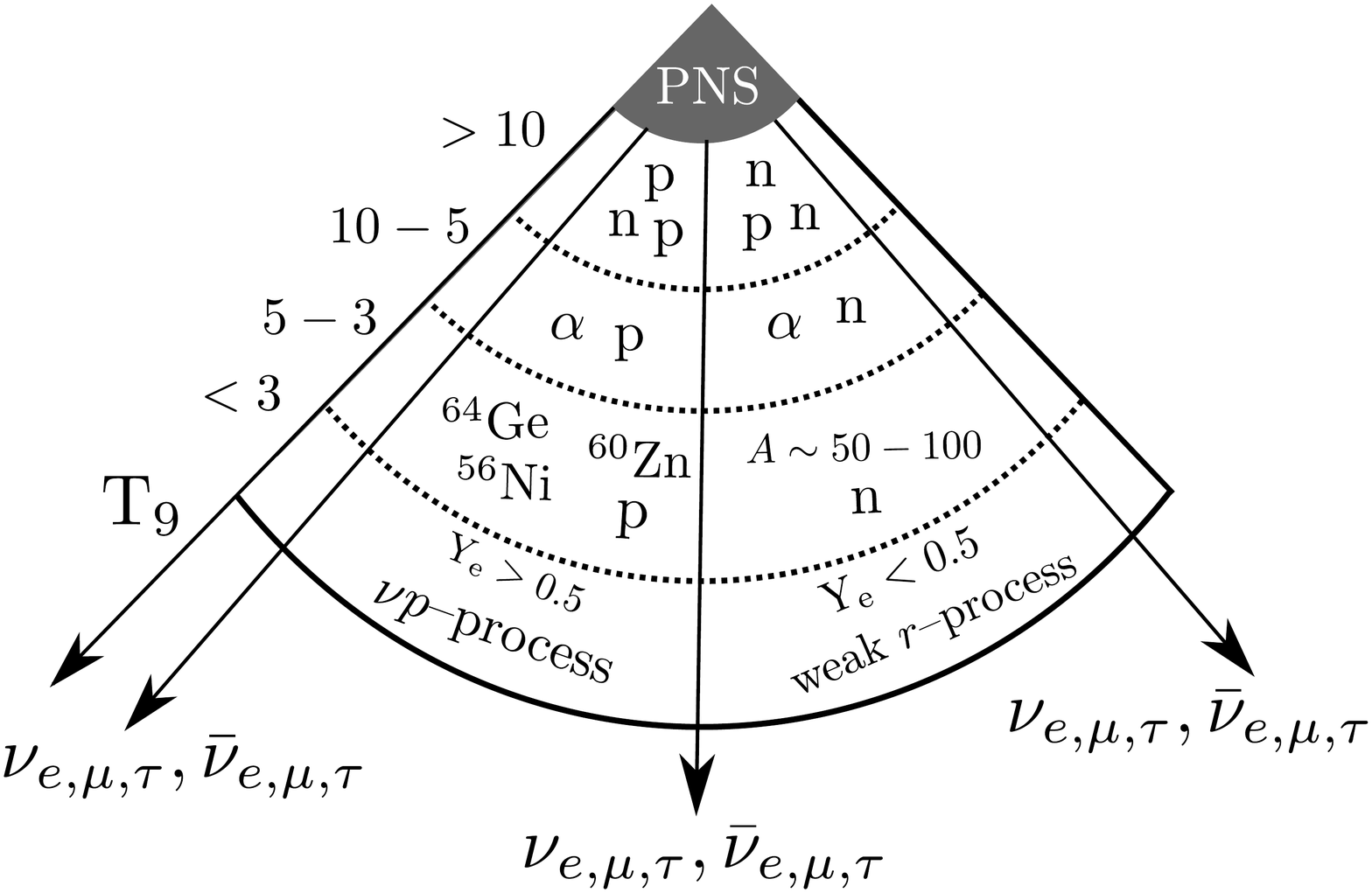}
    \caption{Simplified schematic of the nucleosynthesis in neutrino--driven wind  ejecta. The different stages and outcomes ($\nu p$--process and weak \emph{r}--process) are shown. The figure is adapted from~\citet{jose2011nuclear}.}
\label{fig:1}
\end{figure}

To summarize the $\nu p$--process, the neutrino--driven wind ejects very hot
($T>$ 10~GK) and proton--rich material from the protoneutron star (PNS) (see
Figure~\ref{fig:1}). At these extreme temperatures, the ejecta consist mainly of nucleons
from dissociated nuclei. As the wind expands and cools down, nuclear statistical
equilibrium (NSE) assembles these nucleons into mainly $\isotope[56][]{Ni}$ and $\alpha$
particles (which are synthesised via the hot \emph{pp}--chain
sequence~\citep{wiescher1989hot}) with an excess of free protons. At T$<$ 3--4~GK,
$\isotope[56][]{Ni}$ can rapidly capture free protons. However, the reaction flow cannot
move beyond $\isotope[64][]{Ge}$, which has a relatively long $\beta^+$ half--life of
1.06~min. This issue is resolved by electron antineutrino captures on free protons via the
$p(\bar{\nu}_e,e^+)n$ reaction, which produce a small amount of free neutrons,
$10^{-11}-10^{-12}$ of the total mass. At temperature drops from 3 to 1.5~GK, the much faster
$(n,p)$ reaction on $\isotope[56][]{Ni}$, followed by a sequence of radiative proton captures, \emph{i.e.}
$(p,\gamma)$ reactions, and further $(n,p)$ reactions bypass $\isotope[64][]{Ge}$ and similar waiting--points, such as
$\isotope[68]{Se}$ and $\isotope[72]{Kr}$ with half--lives of 35.5~s and 17.1~s,
respectively. The reaction flow follows the $Z = N$ line up to the molybdenum region and
then moves into more neutron--rich isotopes ($Z < N$) between molybdenum and tin.
Finally, as the temperature drops below $T< 1.5$~GK, $(p,\gamma)$ reactions freeze--out
due to the Coulomb barrier, and the produced nuclei decay back to stability,
with $\isotope[56]{Ni}$ still being the most abundant
nucleus in the plasma.

%Uncertainties of νp-process
The uncertainties of $\nu p$--process nucleosynthesis, mainly attributed to the supernova
dynamics and the underlying nuclear physics input, have been explored by many
groups~\cite{wanajo2011uncertainties, arcones2011production, arcones2012impact, frohlich2012reaction, jacobimsc, nishimura2019uncertainties} since it was first proposed.
The most crucial component for a successful $\nu p$--process is the electron fraction
$Y_e$ of the ejecta. Recent hydrodynamical studies with proper neutrino transport
have shown that $Y_e$ can lie between 0.5 and 0.6 before the onset of $\nu p$--processing
at T= 3~GK. Sensitivity studies have explored a variety of $Y_e$ values, ranging from 0.5 to
0.8, and suggest that a higher $Y_e$ leads to a more efficient $\nu p$--process (\textit{i.e.} production of heavier nuclei).

Concerning the nuclear physics input of the $\nu p$--process the main uncertainties
arise from a handful of reactions, and the nuclear masses along the reaction path.
The two most important reactions that dominate the nucleosynthesis in this scenario
are the bottleneck $\isotope[56][]{Ni}(n,p)\isotope[56]{Co}$ and triple--$\alpha$
-- $\isotope[4][]{He}(\alpha\alpha,\gamma)\isotope[12][]{C}$ -- reactions.
The former is always the first step of the $\nu p$--process and as a result controls
the reaction flow, with a smaller rate yielding a more efficient nucleosynthesis, since
the free neutrons synthesized from neutrino captures are captured by nuclei with
$30 \leq Z \leq 42$, acting as ``neutron poisons''.

The triple--$\alpha$ reaction controls the production of $\alpha$--particles, protons and
the $\isotope[56]{Ni}$ seed before the onset and during the $\nu p$--process.
Therefore it controls completely the neutron--to--seed ratio
$\Delta_n$, as defined by~\citet{pruet2006nucleosynthesis}.
Our current knowledge of this reaction, despite its importance, is still
limited and bears large experimental uncertainties. The three rates that are
most commonly used in nucleosynthesis studies are those
from References~\cite{caughlan1988thermonuclear, angulo1999compilation, fynbo2005revised}.
In addition,~\citet{jin2020enhanced} recently showed that an enhanced
triple--$\alpha$ reaction, due to an in--medium width change of the Hoyle state,
suppresses the production of \textit{p}--nuclei in the $\nu p$--process.

In the sensitivity study of~\citet*{wanajo2011uncertainties}, some
alternative pathways were explored. In particular, the authors found that there are
a couple of two--body reaction sequences, namely
$\isotope[7][]{Be}(\alpha,\gamma)\isotope[11][]{C}(\alpha,p)\isotope[14][]{N}$
and $\isotope[7][]{Be}(\alpha,p)\isotope[10][]{B}(\alpha,p)\isotope[13][]{C}$,
which compete with the triple--$\alpha$ reaction, the main link between the
\textit{pp}--chain ($A <12$) and CNO ($A \geq 20$) region, at the relevant temperature
region $T= 1.5-3$~GK. This competition affects the $\Delta_n$ factor and as a result, the reaction flow
and the final elemental abundances. The authors studied the sensitivity of the
final abundances by multiplying and dividing the \beag~reaction rate by factors of 2 and 10. This rate variation affected the production of light \textit{p}--nuclei with
$90 < A <110$ up to an order of magnitude. A faster \beag~rate leads to increased production of intermediate--mass
nuclei that remove protons from the environment, acting as ``proton poisons''. Subsequent
studies, such as that by~\citet{nishimura2019uncertainties}, also acknowledge the
importance of the \beag~reaction, but do not provide a  quantitative impact in the
production of \textit{p}--nuclei. As we shall discuss in detail in Section~\ref{sec:2}, the
\beag~reaction rate is not well known in the relevant temperature range due to unknown
resonance strengths, and thus an experimental study is required.

In the present work we report on the first inverse kinematics study of
the \beag~reaction, using the DRAGON recoil separator and an intense
$\isotope[7][]{Be}$ radioactive ion beam from ISAC. The paper is structured as follows: in Section~\ref{sec:2} we discuss the
previous measurements regarding resonances of the
\beag~reaction. In
Sections~\ref{sec:3} and~\ref{sec:4} we present the experimental details of the
present work along with the analysis procedures, and finally we discuss our results and conclusions in Sections~\ref{sec:5} and~\ref{sec:6}.

\section{Previous Measurements}
\label{sec:2}

Our current understanding of the \beag~
reaction over the energy region relevant to $\nu p$--process
nucleosynthesis is based on three experimental
studies~\cite{hardie1984resonant, wiescher1983c, yamaguchi2013alpha}.

\begin{figure}[ht!]
    \centering
    \includegraphics[width=.55\textwidth]{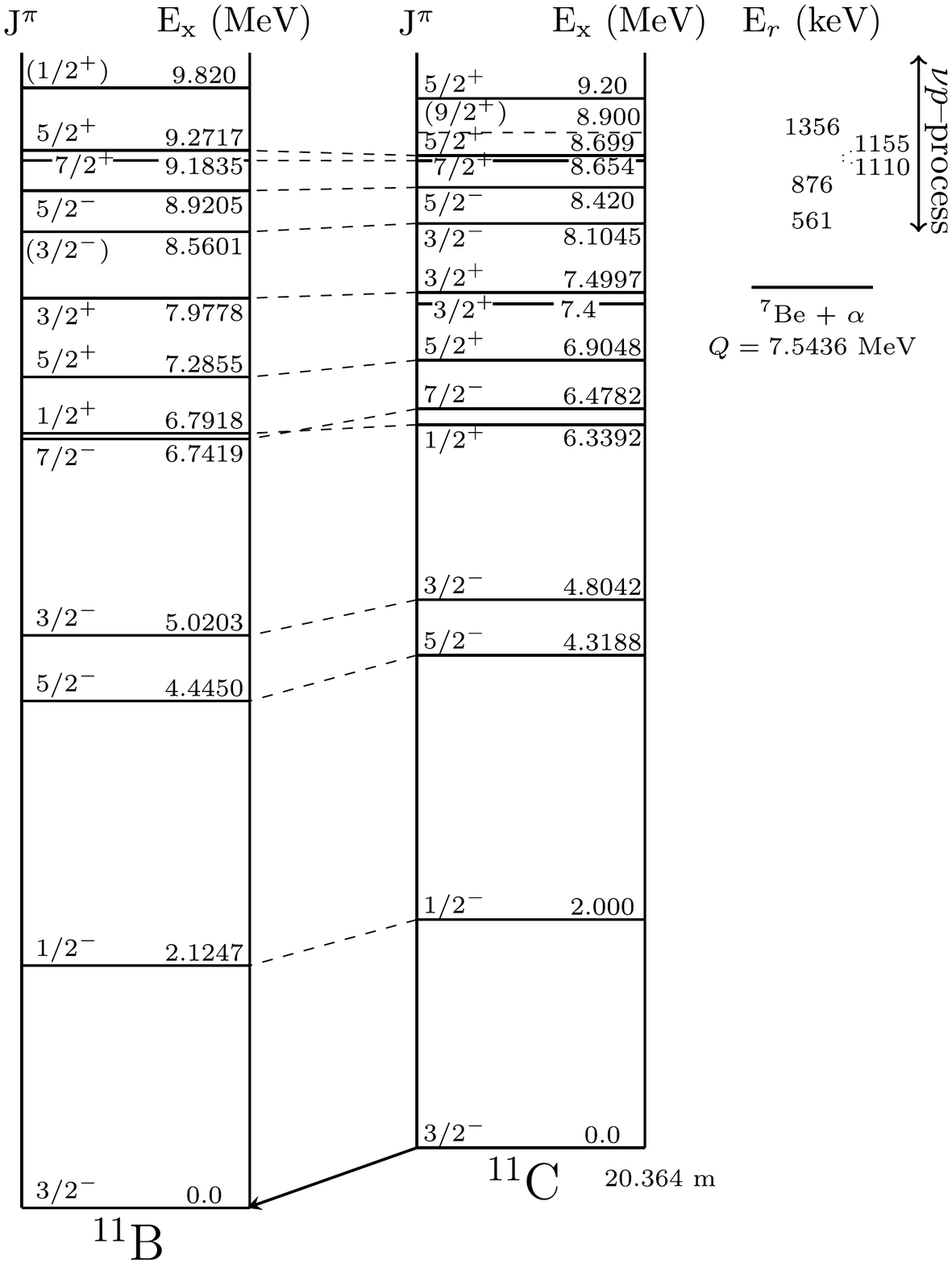}
    \caption{Partial level scheme of the mirror nuclei $\isotope[11][]{C}$
    and $\isotope[11][]{B}$ adopted from~\citet{kelley2012energy}, with the addition of the $E_x = 8.900$~MeV state from~\citet{yamaguchi2013alpha}. The dashed lines indicate isobaric analog states, and next to the $\isotope[11][]{C}$ scheme we present the $\alpha$ separation energy $Q_\alpha$, the resonances of the \beag~ reaction, $E_r$, in keV and the relevant energy region for $\nu p$--process nucleosynthesis.}
\label{fig:2}
\end{figure}

Figure~\ref{fig:2} shows the current level structure of $\isotope[11][]{C}$ along with its
mirror $\isotope[11][]{B}$, and in Table~\ref{tab:1} we summarize the resonance parameters for the
\beag~ reaction from the $A=11$ evaluation of~\citet{kelley2012energy}.

%Hardie
The two lowest--lying energy resonances of the
\beag~
reaction, which correspond to the $E_x$ = 8.105 and 8.420~MeV levels in
$\isotope[11][]{C}$, were studied by~\citet{hardie1984resonant} in
forward kinematics at Argonne National Laboratory. The authors used two methods to calculate
the resonance strengths: the first was the thick target yield formula (similar to Equation~\ref{eq:3}), and the second was a
complementary relative method which employed the presence of $\isotope[7][]{Li}$ in the
target, and the fact that they were studying the $\isotope[7][]{Li}(\alpha,\gamma)\isotope[11][]{B}$
reaction in the same campaign. More specifically, the relative method provided
the resonance strength ratio between the resonances of interest in
$\isotope[7][]{Be}(\alpha,\gamma)\isotope[11][]{C}$ and the known 660~keV
($E_x$ = 9.272~MeV in $\isotope[11][]{B}$) resonance of
the $\isotope[7][]{Li}(\alpha,\gamma)\isotope[11][]{B}$ reaction, reported in the same work.
The main advantage of this method is that both the $\isotope[7][]{Li}$:$\isotope[7][]{Be}$
ratio in the target and the detector efficiencies are more accurately known than the number
of $\isotope[7][]{Be}$ atoms alone and the absolute efficiencies. Nevertheless, one has to
include an extra uncertainty factor from the
$\isotope[7][]{Li}(\alpha,\gamma)\isotope[11][]{B}$ resonance. The adopted values for the two resonance strengths in Table~\ref{tab:1} are the weighted averages of the two methods.

%Wiescher
\citet{wiescher1983c} studied the $E_x$ = 8.654 and 8.699~MeV levels in
$\isotope[11][]{C}$, which correspond to the
1110 and 1155 keV\footnote{All resonance energies are expressed in the center of mass system.}
resonances of the \beag~reaction. They used
the $\isotope[10][]{B}(p,\gamma)\isotope[11][]{C}$ reaction in
forward kinematics employing three different
linear accelerators, covering a wide energy range ($E_x$ =  8 -- 10.7 MeV). In all
three experimental setups, several detectors were used, allowing for angular
distribution measurements. The authors observed primary $\gamma$ transitions from the
$E_x$ = 8.654 and 8.699~MeV states and calculated the ratio $\Gamma_\gamma/\Gamma$
for them using the cross sections from the $\gamma$ ray and $\alpha$--particle channels,
$\sigma(p,\gamma)/\sigma(p,\alpha)$.

\begin{table*}[ht!]
\centering
\caption{
Resonance parameters for the \beag~ reaction from~\citet{kelley2012energy}. The parameters for the resonance noted with a $\diamond$ are adopted from~\citet{yamaguchi2013alpha}. Tentative assignments and estimates are presented in parentheses. The resonances noted with a $\dagger$ were studied in the present work. Values noted with a $\Vert$ were adopted from the mirror nucleus $\isotope[11][]{B}$. The proton partial widths $\Gamma_p$ have been calculated using $C^2S=1$.}
\begin{tabular}{cccccccc}
\hline \hline
$E_x$~(MeV) & $E_r$~(keV) & $J^\pi$ & $\Gamma_\alpha$ & $\Gamma_\gamma$ &   $\Gamma_p$ & $\ell$ & $\omega\gamma$ (eV)\\ \hline
7.4997(15) & -43.9(15) & $3/2^+$  & 2.2(1.6)~eV & 1.15~eV$^{\Vert}$ & - & 1 & $\cdots$\\
8.1045(17) & 560.5(17) & $3/2^-$ & 6$^{+12}_{-6}$~eV & 0.350(56)~eV & - & 0 & 0.331(41)\\
8.420(2)$\dagger$ & 876(2) & $5/2^+$  & 12.6(38)~eV & 3.1(13)~eV   & - & 2 & 3.80(57) \\
8.654(4)$\dagger$ & 1110(4) & $7/2^+$  & $\leq 5$~keV  &    & - & 3 & $\cdots$\\
8.699(2)$\dagger$ & 1155(2) & $5/2^+$  & 15(1)~keV & 1.15(16)~eV$^{\Vert}$ & - & 1 & $\cdots$\\
8.900$^\diamond$    & 1356       & ($9/2^+$) & $>$8 keV    &  ~ & - &   (3) &  (1.2)   \\
9.645(50) & 2101(50) & ($3/2^-)$   & 210(40)~keV & 17~eV$^{\Vert}$ & 48(9)~eV  & 0 & $\cdots$\\
9.780(50) & 2236(50) & $(5/2^-)$  & 240(50)~keV  & 1~eV$^{\Vert}$ &  520(100)~eV & 2 & $\cdots$\\
9.970(50) & 2426(50) & $(7/2^-)$   & 120(20)~keV  & 1~eV$^{\Vert}$ & 760(140)~eV  & 2 & $\cdots$\\
10.083(5) & 2539(5) & $9/2^+$  & $\approx$230~keV  & $<$0.2~eV$^{\Vert}$ & 900(180)~eV & 3 & $\cdots$ \\
\hline \hline
\end{tabular}
\label{tab:1}
\end{table*}

%Yamaguchi-san
The most recent study relevant to the $\isotope[7][]{Be}(\alpha,\gamma)\isotope[11][]{C}$
reaction was performed by~\citet{yamaguchi2013alpha} using the
low--energy radioactive ion beam facility CRIB~\cite{yanagisawa2005low} at CNS in RIKEN, Japan.
The $\isotope[7][]{Be} + \alpha$ resonant scattering and
$\isotope[7][]{Be}(\alpha,p)$ reaction measurements were performed using the
thick--target method in inverse kinematics and provided the excitation functions for
$E_x$ = 8.7--13.0 MeV. The \emph{R}--matrix analysis of the data shows two small
peaks in the low energy region, between 8.90 and 9.20~MeV. The first one is considered to be
the known $5/2^+$ state at 9.20~MeV observed by~\citet{wiescher1983c}. The second one, located at around 8.90~MeV,
is regarded by the authors as a new resonance. However, they argue that this spectral
feature could also originate from either the $E_x=$~8.655 or the 8.699~MeV states since their the energy uncertainty was quite large in this energy region. Finally, all this information about the \beag~ resonances is summarised in Table~\ref{tab:1}.

%Current rate
The current \beag~ reaction rate is based
on a calculation from NACRE (I and II)~\cite{angulo1999compilation, xu2013nacre} and includes
contributions only from the 561 and 876 keV resonances, for which
experimentally measured strengths exist, and the non--resonant (DC) contribution is adopted
with the same parameters as those of the mirror
$\isotope[7][]{Li}(\alpha,\gamma)\isotope[11][]{B}$ reaction, for
T $<$ 0.7~GK. Contributions from the broad resonances at 2101, 2236, 2426~\&~2539~keV were also included in NACRE--II and affect the reaction rate for T $>$ 2~GK. In NACRE--I the same high energy part of the reaction rate was estimated using Hauser--Feshbach calculations. The rate that was used  in the sensitivity study by~\citet{wanajo2011uncertainties} was the one from~\citet{angulo1999compilation} (NACRE--I), which is uncertain by factors of 1.87--2.54 in the relevant temperature region. The NACRE--II reaction rate is uncertain
by factors of 1.76–1.91 for T = 1.5–3 GK. The uncertainties are derived from
their Potential Model (PM), which is used to reproduce the experimental astrophysical $S$--factor data. $S(E) \equiv (E/e^{-2\pi \eta})~\sigma(E)$, where $E$ is the center of mass energy, and $\eta$ is the Sommerfeld parameter, which is related to the charges and velocities of the interacting particles. More specifically, the uncertainties are calculated by using the maximum and minimum parameters of the PM.
It is also worth noting that the sub--threshold resonance at $E_x$ = 7.50~MeV ($E_r= -43.9$~keV) has a large
contribution at low temperatures, below T $\approx$ 0.3~GK and according to~\citet{descouvemont19957be},
it could affect the production of $\isotope[7][]{Li}$ (fed by the decay of $\isotope[7][]{Be}$) in classical novae~\cite{hernanz1996synthesis}.

\section{Experimental Details}
\label{sec:3}

The measurements of this work were carried out using the DRAGON recoil
separator~\cite{hutcheon2003dragon} at TRIUMF, Canada's particle accelerator
centre in Vancouver, BC. Intense beams of $\isotope[7][]{Be}^{+}$ were produced using
the ISOL technique, by bombarding thick ZrC and graphite targets with 55~$\mu A$
500~MeV protons from the {TRIUMF} cyclotron. The A= 7 isobars, mainly $\isotope[7][]{Li}$ and $\isotope[7][]{Be}$, were extracted
from the target through a high-resolution mass separator, and the beryllium ionization was enhanced using the TRIUMF Resonant Ionization Laser Ion Source
(TRILIS)~\citep{lassen2005resonant}. After the ion source, the beam was
transported through the ISAC high resolution (M/$\Delta$M  = 2000) mass separator
and then accelerated through the ISAC--I Radio--Frequency Quadrupole
(RFQ) and Drift--Tube Linac (DTL) to energies
so that each resonance was centered in the gas
target (see Table~\ref{tab:2} for details).
The beam energies were chosen in order to
cover center--of--mass windows of $1157 \pm 24$~keV , $1111 \pm 13$~keV,
and  $878 \pm 17$~keV, across the gas target volume.
To ensure a pure, contaminant--free radioactive ion beam,
an additional carbon stripping foil of 20~$\mu g$/cm$^2$ was placed
downstream of the DTL allowing fully stripped $\isotope[7][]{Be^{4+}}$ to
be selected using a bending magnet for transport to DRAGON, thus
eliminating the main isobaric contaminant $\isotope[7][]{Li}$.
This technique has also been used in other radioactive beam facilities~\cite{GAELENS200348}.
Finally, $\isotope[7][]{Be^{4+}}$ was delivered at the helium--filled windowless gas target
of DRAGON at mean intensities of $\sim 1.3-5.8 \times 10^8$~pps (see Section~\ref{sec:beamnorm} for normalisation details of the intensities).

\begin{table}[ht!]
\centering
\caption{Beam and gas target properties for the two independent measurements of
the present study\footnote{The 1110 keV resonance was studied in two independent
measurements, due to a low recoil yield in the first measurement.}.}
\begin{tabular}{ccccc}
\hline \hline
 &  E\textsubscript{beam} (\textit{A} keV)  & E\textsubscript{lab} (MeV) &  I\textsubscript{beam} & P\textsubscript{target} (Torr)    \\
~ & ~ & ~ & $\times 10^8$ (s\textsuperscript{-1}) & ~ \\ \hline
Run 1 & 464.2(3) & 3.249(2) & 1.33(7)     &  7.9(1)   \\
Run 1 & 442.7(2) & 3.099(1) & 2.06(8)     &  5.06(6)  \\ \hline
Run 2 & 441.8(2) & 3.093(1) & 5.83(2)     &  4.89(3)  \\
Run 2 & 351.8(3) & 2.463(2) & 3.45(12)    &  5.75(4)  \\
\hline \hline
\end{tabular}
\label{tab:2}
\end{table}

\begin{figure}[ht!]
    \centering
    \includegraphics[width=.5\textwidth]{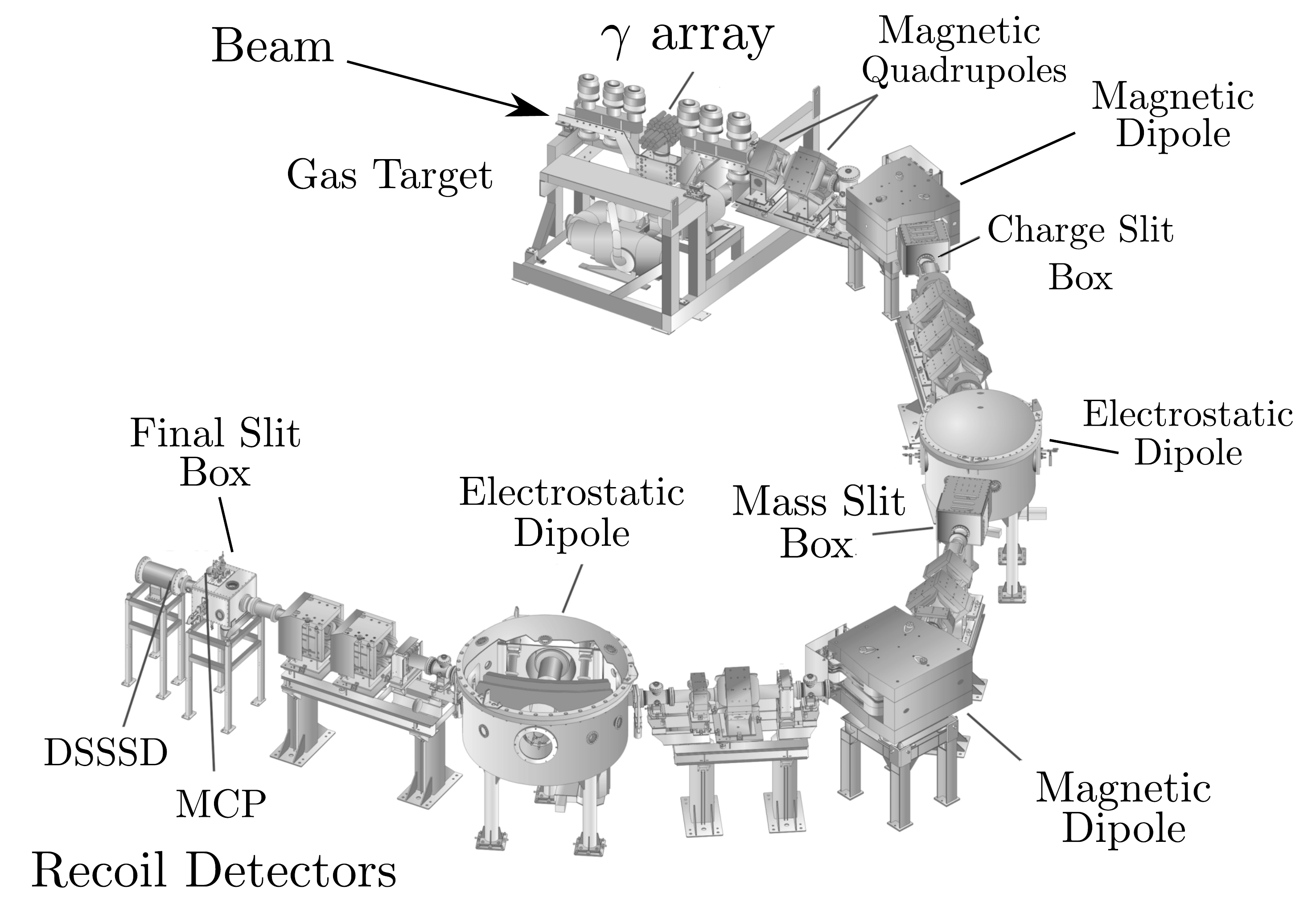}
    \caption{Schematic representation of the DRAGON recoil separator. The main components are shown.}
\label{fig:3}
\end{figure}

% General Description of DRAGON
DRAGON has four main components: (a) the windowless, differentially pumped,
recirculating gas target, (b) the $\gamma$ ray detector array, (c) the
electromagnetic mass separator and (d) the recoil detection system,
which are shown in the schematic of Figure~\ref{fig:3}.

% BGO
The $\gamma$ ray array consists of 30 BGO scintillator crystals with photo--multiplier
tubes (PMTs) covering $89-92\%$ of the 4$\pi$ solid angle~\cite{ruiz2014recoil}.
The segmented array allows for the detection of individual prompt $\gamma$ rays from
the radiative capture reactions inside the gas target and the tagging of the associated
recoil particles, which provides an additional background reduction in the focal plane
detectors.

% Main separator
The DRAGON electromagnetic mass separator consists of two magnetic (M) and two electric
dipoles (E) in a MEME configuration. The two--stage separation begins with the first
magnetic dipole (MD1), which selects a single charge stage to be transmitted through DRAGON. For our study, we tuned DRAGON to the $q= 2^+$ charge state
for all resonances.
Recoils that do not have the aforementioned charge are deflected to slits that are
located downstream of the MD1. Subsequently, the recoils are led to the first
electric dipole (ED1) where they are separated according to mass. ED1 is followed by the
second stage of magnetic and electric dipoles (MD2 and ED2), until the beam reaches the
focal plane, where the heavy ion detectors are located.

% End detectors and DAQ
Close to the focal plane of DRAGON we employed a microchannel plate (MCP)
and a double--sided silicon--strip detector (DSSSD). The MCP provided the starting timing
signal for a local time--of--flight (TOF) transmission
measurement~\cite{vockenhuber2009transmission}. In the DSSSD, the recoils are stopped,
their kinetic energy and position are measured and the stopping timing signal is recorded.
In addition, we employed the prompt $\gamma$ rays from the BGO array as a starting signal
for a ``separator TOF'' measurement for the coincidence analysis (see
Section~\ref{sec:pid}). The data were recorded using a state--of--the--art
time--stamp--based data acquisition system (DAQ)~\cite{christian2014design}.

It is worth mentioning that such a measurement using DRAGON,
and any other recoil separator dedicated to resonance strength measurements,
is quite challenging, due to geometric considerations. The maximum recoil angle of the reaction at the resonance energies of interest can be calculated using the following equation,
\begin{equation}
    \theta_{r,max} = \arctan\left( \frac{E+Q}{\sqrt{2 m_1 c^2 \left(\frac{m_1+m_2}{m_2}\right) E}} \right)
\end{equation}
where $E$ is the center of mass energy, $Q$ is the reaction $Q$ value and $m_1,m_2$ the masses of the projectile and the target nuclei, respectively. This corresponds to a single
$\gamma$ ray emission to the recoil nucleus ground state at $90^{\circ}$ in the center--of--mass system.

For the \beag~ reaction maximum recoil angles are
$\theta_{r,max}$ = 42.67~mrad for the 1155~keV, 43.3~mrad for the 1110~keV and 47.42~mrad for the 876~keV. These numbers far exceed the nominal angular acceptance of DRAGON,
$\mathrm{\theta_{DRAGON}}= \pm 21$~mrad). For this reason, we performed detailed
{\scshape Geant} simulations of DRAGON~\cite{gigliotti2003calibration, gigliotti2004efficiency}
to extract the transmission of the recoils though the separator ($\eta_{separator}$), and
in addition, the BGO array efficiency ($\eta_{BGO}$), which are used for the resonance strength calculations.
References~\cite{ruiz2014recoil,psaltis6li, psaltis2020radiative} provide an in--depth discussion
about this approach, and in Section~\ref{sec:geant} we provide the specifics for the study of the \beag~reaction.

\section{Data Analysis}
\label{sec:4}

We performed yield measurements for three beam energies, corresponding to the
1155, 1110 and 876~keV resonances of the \beag~reaction (see Figure~\ref{fig:2}).
As we have already discussed in Section~\ref{sec:2}, the 876~keV resonance strength has been
measured by~\citet{hardie1984resonant}, while the latter two resonances have unknown
strengths. Our reasoning to re--measure that resonance is two--fold: on the one hand it
is believed to have the greatest impact on the current reaction rate at $\nu p$--process
energies~\cite{xu2013nacre} and on the other hand it will provide one additional
demonstration that DRAGON can measure resonance strengths for reactions in which the
angular spread of the recoils exceeds its nominal
acceptance~\cite{psaltis6li, psaltis2020radiative}.

\subsection{Thick target yield and resonance strength}
\label{sec:yield}

The calculation of thermonuclear reaction rates in a laboratory setting requires the
determination of the reaction cross section. Instead, what is actually measured in
experimental studies is the reaction yield, which can be simply expressed as:

\begin{equation}
    Y = \frac{\mathcal{N_R}}{\mathcal{N_B}}
\end{equation}
where $\mathcal{N_R}$ is the number of reactions that occur and $\mathcal{N_B}$ is the
number of incident beam particles. In fact, an experimental setup has a finite detection
efficiency, in our case $\eta_{DRAGON}$, meaning that it does not detect the total
number of reactions, but rather a fraction of it, $N_r$. According to the analysis mode
that we use, singles or coincidences (see Section~\ref{sec:pid}), $\eta_{DRAGON}$ can be
either $\eta_{DRAGON}^{singles}= \eta_{separator}~f_q ~\eta_{MCP}~ \eta_{DSSSD}~
\eta_{live}^{singles}$, or $\eta_{DRAGON}^{coinc}= \eta_{separator}~ f_q~ \eta_{BGO}~ \eta_{MCP}~ \eta_{DSSSD}~ \eta_{live}^{coinc}$, respectively.
The experimental yield is then given by:

\begin{equation} \label{eq:2}
    Y = \frac{\mathcal{N}_r}{\mathcal{N_B}~\eta_{DRAGON}}
\end{equation}

We can also express the energy--dependent reaction yield as a relation between the cross
section $\sigma(E)$ and the target thickness $\Delta E$, or better, the stopping power of the target $\varepsilon(E)$, for beam energy $\mathrm{E_{beam}}$ using:
\begin{equation}\label{eq:3}
    Y(\mathrm{E_{beam}}) = \int_{E-\Delta E}^{E} \frac{\sigma(E)}{\varepsilon(E)} dE
\end{equation}
For narrow resonances with constant stopping power over the resonance width, which can be found in reactions relevant for astrophysics, we can calculate the integral of Equation~\ref{eq:3} analytically using a single--level Breit--Wigner (Lorentzian) cross section
profile~\cite{fowler1948gamma}. Specifically, in the case of an infinitely thick target,
that is $\Delta E \rightarrow \infty$, or equivalently $\Delta E \gg \Gamma$, we have:

\begin{equation}
    Y(\mathrm{E_{beam}}) = \frac{\lambda^2_r}{2\pi} \frac{\omega \gamma}{\varepsilon_r} \left[ \tan^{-1}\left( \frac{\mathrm{E_{beam}}-E_r}{\Gamma/2}\right) + \frac{\pi}{2} \right]
\end{equation}

where $\lambda_r$ and $\varepsilon_r$ are the de Broglie wavelength and the target stopping
power in the center of mass system, $E_r$ is the energy of the resonance, $\Gamma$ is
its width and $\Delta E$ is the target thickness. Solving for $\mathrm{E_{beam}}=E_r$, we can obtain a
simple expression for the reaction yield and the resonance strength $\omega \gamma$:

\begin{equation}
    \omega \gamma = \frac{2 Y_{\Delta E \rightarrow \infty} \varepsilon}{\lambda_r^2} \frac{m_1}{m_1+m_2}
\end{equation}
where the reaction yield $Y_{\Delta E \rightarrow \infty}$ is given by
Equation~\ref{eq:2} and $\varepsilon$ is the target stopping power in the laboratory frame
-- a discussion on how it is measured in DRAGON experiments can be found in Section~\ref{sec:dedx}.

\subsection{Particle Identification}
\label{sec:pid}

\begin{figure}[ht!]
    \centering
    \includegraphics[width=.5\textwidth]{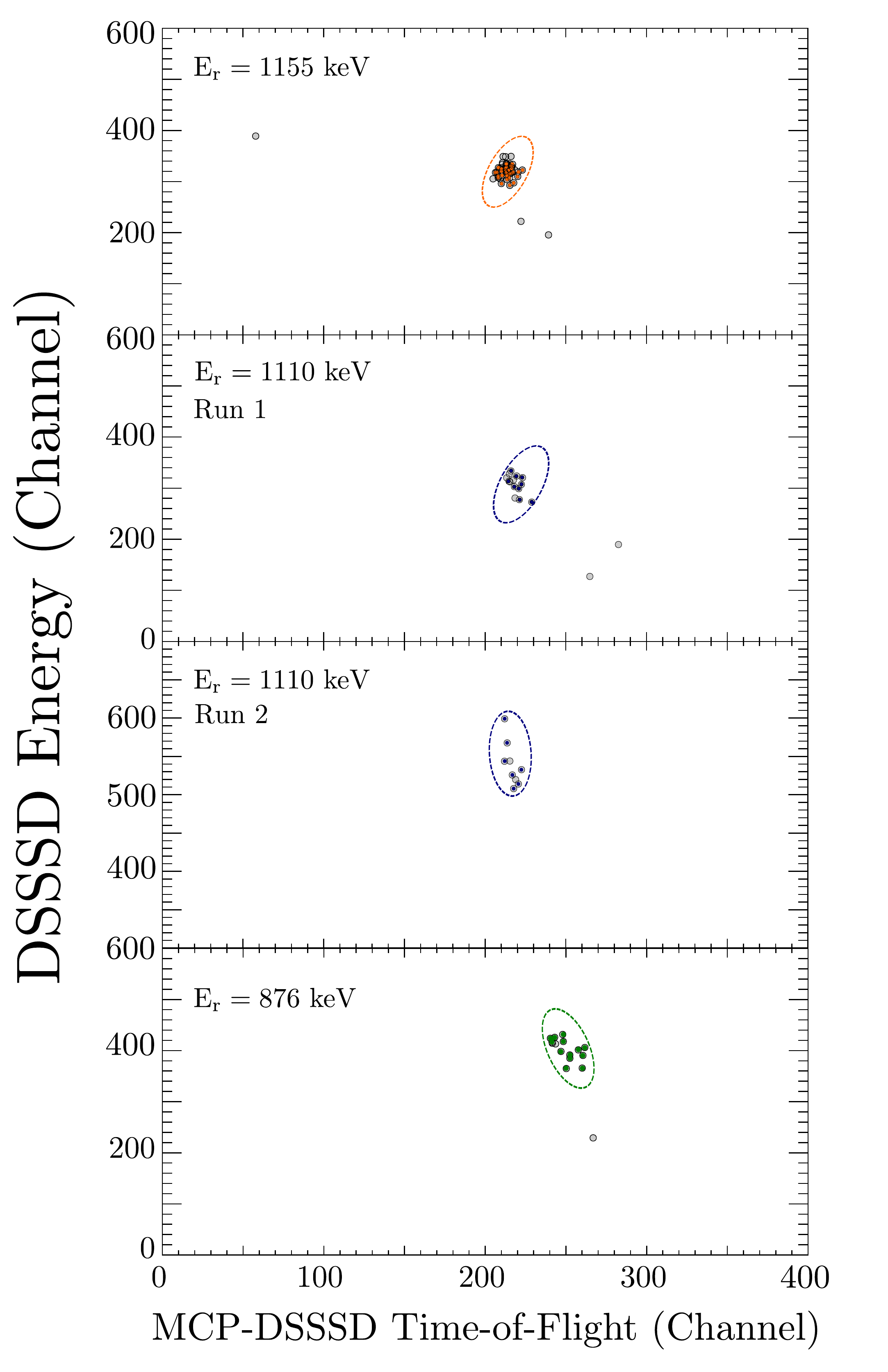}
    \caption{Particle identification plot for the $\isotope[11][]{C}$ recoils for each
    of the resonances we studied in the present work using the
    local Time--Of--Flight transmission measurement and the energy deposited in the DSSSD at the focal plane of DRAGON. The coloured and grey points correspond to coincident and singles recoils events, respectively. For the 1110~keV, we show the two independent measurements in separate panels. See the text for details.}
\label{fig:4}
\end{figure}

The first step towards determining the reaction yield and subsequently the strength of a resonance is the identification of the reaction products or recoils. For this, we employed two distinct
methods: a detection of $\isotope[11][]{C}$ recoils in singles, using the DSSSD and a local
TOF (MCP--DSSSD), and in $\gamma$--recoil coincidences using the separator TOF (BGO--DSSSD).
Figure~\ref{fig:4} shows typical particle identification plots for the three resonances
of the \beag~reaction both in singles (grey points) and in coincidences (coloured points), with additional software
cuts, such as the energy range of the $\gamma$ rays in the BGO array and the energy deposited in the DSSSD, providing further recoil discrimination. It is evident that for the 876 and 1110~keV
resonances the yield is low, but the signal is clear and without any unwanted background, such
as unreacted, ``leaky'' beam. This is consistent with the fact that DRAGON is able to reject unreacted beam particles from ($\alpha,\gamma$)
reactions very efficiently and has demonstrated a beam suppression of $>10^{13}$~\cite{sjue2013beam}. In the present experiment, the rejection is higher, due to the use of a fully--stripped beam and the fact that the selected carbon recoils have a very different charge state ($4^+$ vs $2^+$).

\begin{figure}[ht!]
    \centering
    \includegraphics[width=.5\textwidth]{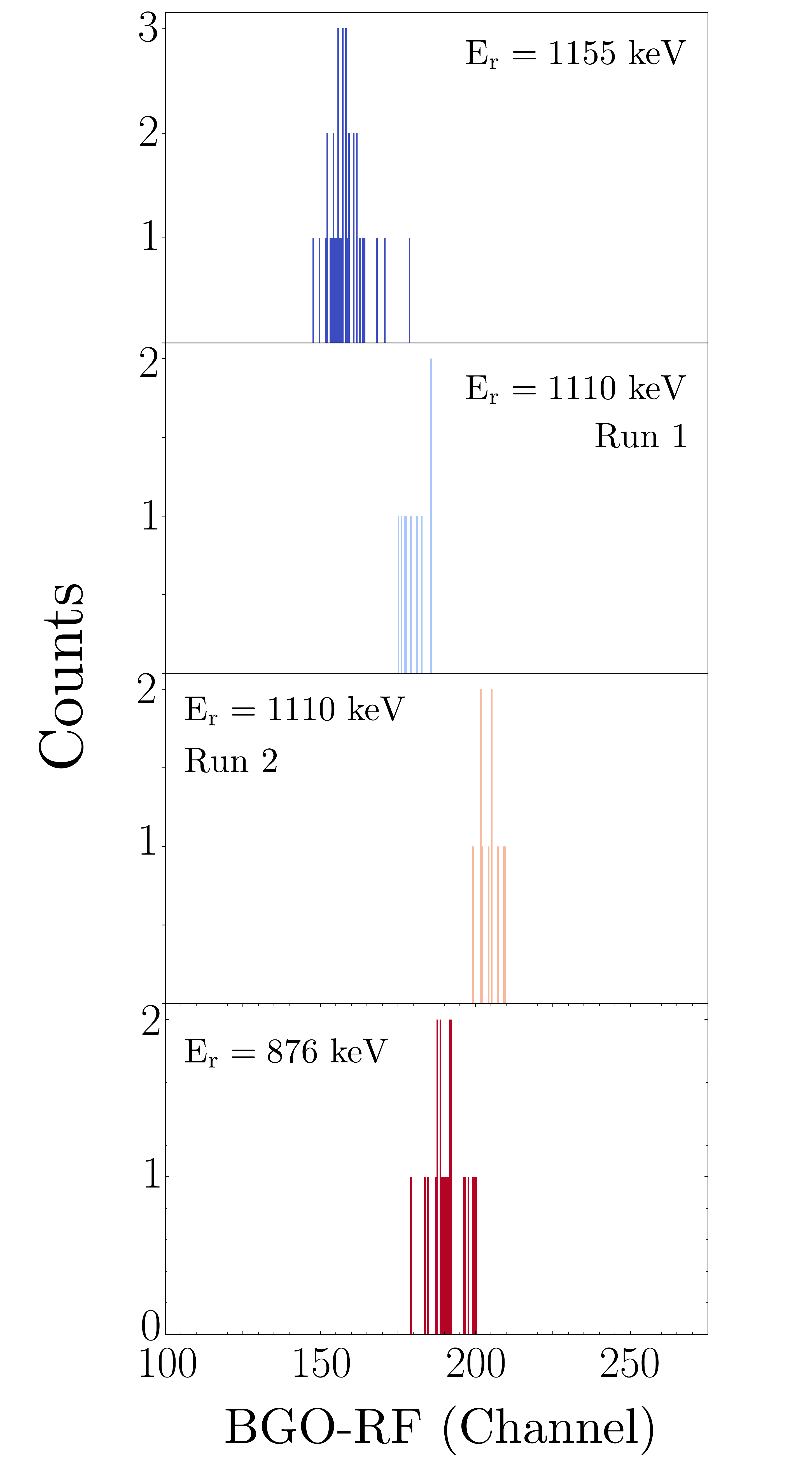}
    \caption{Particle identification plot using the BGO versus accelerator RF Time--of--Flight for coincident recoil events. The RF period is $\tau_{RF}=$ 84.8~ns.}
\label{fig:5}
\end{figure}

In addition to the aforementioned methods, we could also identify the recoils of interest using a timing
signal of the 11.8 MHz ISAC-I radio frequency quadrupole (RF) accelerator and
the capture of a coincidence $\gamma$ ray by the BGO array
(BGO--RF)~\cite{hutcheon2003dragon, christian2014design}.
Figure~\ref{fig:5} shows the results for each resonance.

\subsection{Beam Normalization}
\label{sec:beamnorm}

We monitored the $\isotope[7][]{Be}$ beam current throughout the experiment
using silicon surface barrier detectors (SSB) at well defined laboratory
angles of 30$^\circ$ and 57$^\circ$ with respect to the beam axis by
detecting the elastically scattered target particles. Due to the low count rate
in the SSB detectors, we did not use SSB measurements for short time windows $\Delta t$,
before and after each yield measurement to calculate the beam normalization $\mathcal{R}$
factor, as is typical in DRAGON experiments~(see for example the works in Refrences
\cite{hutcheon2003dragon, hager2012measurement, connolly2018direct, PhysRevC.102.035801}). Instead, we first
ensured that the beam current during each run was stable by checking the  current on the
charge slits after the first magnetic dipole and used the total integrated counts in the SSBs per
yield run to calculate the $\mathcal{R}$ factor, which is given by:

\begin{equation}\label{eq:norm}
\mathcal{R} = \frac{I}{|q \cdot e|} \frac{\Delta t }{N_\alpha } \frac{P}{E_b^2} \eta_{target}
\end{equation}
where $I$ is the average current reading at the upstream Faraday Cup before the gas target, $q$ is the beam charge state ($4^+$),
$e$ is the elementary charge (e= $1.6 \times 10^{-19}$ C), $N_{\alpha}$ is the number of
scattered $\alpha$ particles detected by the surface barrier detectors during the yield run time $\Delta t$, $P$ is the gas target pressure in Torr, $E_b$ is the beam energy in keV/u
and $\eta_{target}$ is the transmission through an empty target. We assume only elastic Rutherford scattering for the target particles and the $E_b^2/P$ factor enters Equation~\ref{eq:norm} to make $\mathcal{R}$ invariant to the chosen beam energy and target pressure~\cite{PhysRevC.69.065803}.

The normalized number of
beam particles $N_{beam}$, is then given by:

\begin{equation}\label{eq:norm2}
  N_{beam} = \mathcal{R} N_\alpha \frac{E_b^2}{P}
\end{equation}

Table~\ref{tab:3} shows the $\mathcal{R}$ factor results for all
the yield measurements of the present work. Note that in our two independent experimental runs we used
different SSB gains, threshold settings, and pre-scalers. For this reason, the $N_{\alpha}$ that we use in Equations~\ref{eq:norm} and~\ref{eq:norm2} to extract $\mathrm{N_{beam}}$ are also different.

\begin{table}[ht!]
\centering
\caption{Beam normalization results for the yield measurements of the present work\footnote{During the two independent experimental runs, we used different SSB gains, threshold settings, and pre-scalers. For this reason the absolute value of the SSB rate is not comparable between these different periods.}.}
\begin{tabular}{cccc}
\hline \hline
~ &  E\textsubscript{beam} (A keV) & $\mathcal{R}$-factor & N\textsubscript{beam}  \\
~ &  & $(\isotope[7][]{Be}/\alpha)$(Torr/keV$^2$) & $\times 10^{13}$ ions  \\ \hline
Run 1 & 464.2(3) & $1.15(2) \times 10^{11}$  & 1.07(2)  \\
Run 1 & 442.6(2) & $1.22(2) \times 10^{11}$  & 1.76(5)  \\ \hline
Run 2 & 441.8(2) & $1.74(4) \times 10^{10}$  & 1.53(4)  \\
Run 2 & 351.8(3) & $2.77(6) \times 10^{10}$  &  2.12(4) \\
\hline \hline
\end{tabular}
\label{tab:3}
\end{table}

\subsection{Carbon in helium charge state distribution}
\label{sec:csd}

DRAGON is tuned to select and transport a single charge state to the final focal plane.
For this reason, an accurate knowledge of the recoil charge state distribution (CSD)
is necessary to determine the total reaction yield. Since the recoil nucleus, in our case $\isotope[11][]{C}$, is unstable,
an abundant and stable isotope of the same element is used instead --- $\isotope[12][]{C}$.
The stable ion beam for this measurement was provided from the microwave ion
source (MWIS) of the ISAC Off--Line Ion Source (OLIS)~\cite{jayamanna2014off}.

\begin{figure}[ht!]
    \centering
    \includegraphics[width=.45\textwidth]{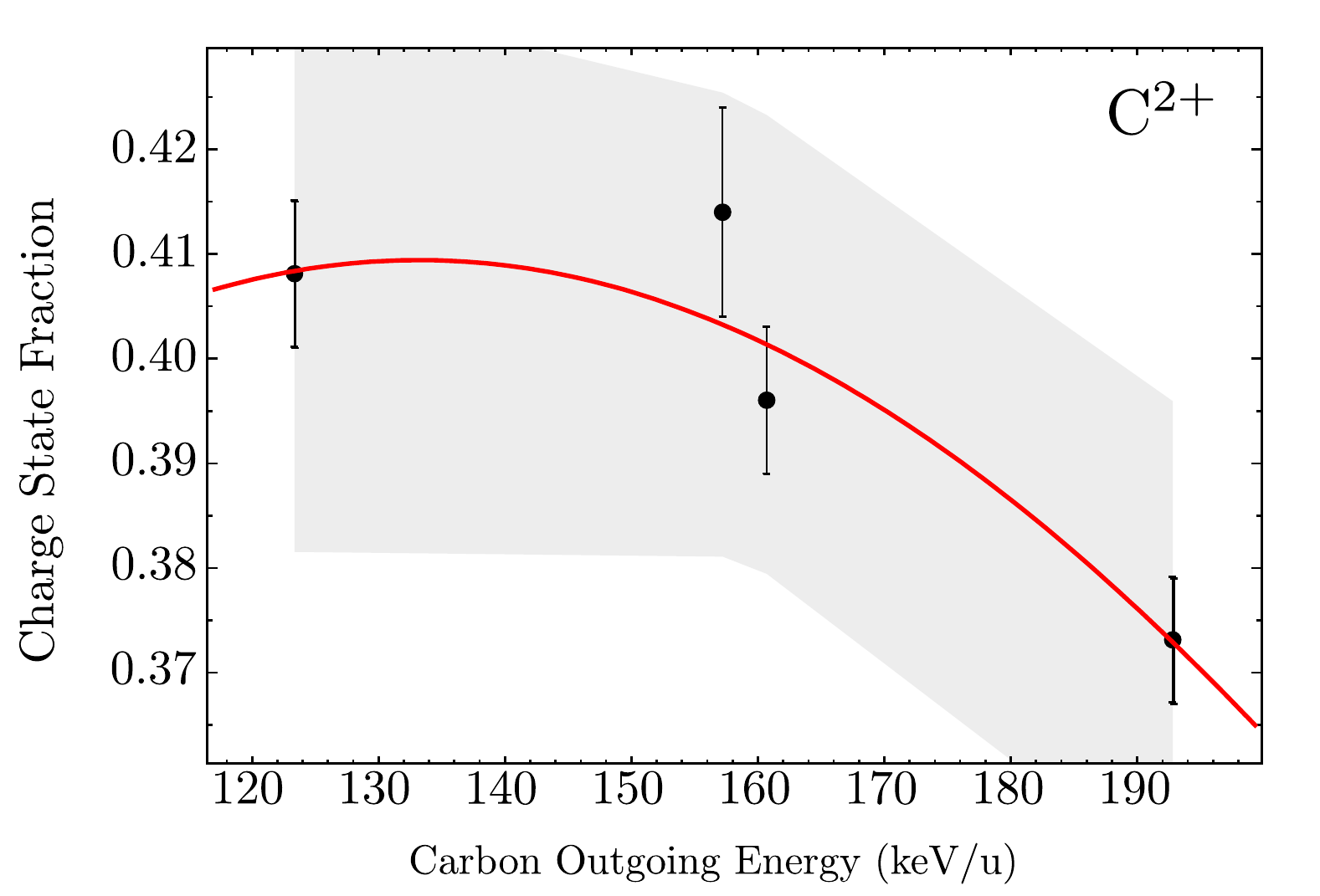}
    \caption{Experimentally measured carbon in helium Charge State Distribution. The fit to the experimental data is a Gaussian function, and the band gives the 1$\sigma$ confidence interval to the fit.}
\label{fig:6}
\end{figure}

At DRAGON, the charge state distributions can be determined experimentally
by measuring the beam current on Faraday cups before
and after the gas target, and comparing it to the current on a
Faraday cup downstream from the first magnetic dipole (see Figure~\ref{fig:3}).
We chose to tune DRAGON to the $2^+$ charge state because according to theoretical calculations~\cite{liu2003charge}, it is the maximum of the distribution, thus providing the highest recoil yield. Figure~\ref{fig:6} shows the results for this charge charge for energies corresponding to the \beag~resonance strengths.

\subsection{$^7$Be stopping power in $^4$He}
\label{sec:dedx}

The stopping power $\epsilon$ of the gas target is an important component for the
calculation of the reaction yield. The advantage of recoil separators, such as DRAGON, is
that the stopping power is measured directly and it is not based on semi--empirical
estimates that can introduce an additional uncertainty factor in the final result.
At DRAGON, the stopping power is measured by varying both the pressure of the gas
target and the magnetic field strength needed to center the beam at a momentum dispersed
angular focus after the first magnetic dipole. Our experimental results
agree to within 6\% with calculations using the {\scshape srim}
code~\citep{ziegler2010srim}, as shown in Table~\ref{tab:4}.

%check this one
\begin{table}[ht!]
\centering
\caption{Summary of the $\isotope[7][]{Be}$ in $\isotope[4][]{He}$ stopping power measurements. The experimental results are compared with the calculations of {\scshape srim}~\citep{ziegler2010srim}. The units of $\epsilon$ are eV/ (10\textsuperscript{15} atoms/ cm$^2$).}
\begin{tabular}{p{0.3\linewidth}p{0.3\linewidth}p{0.3\linewidth}}
\hline \hline
  E\textsubscript{beam} (A keV) &  $\epsilon_{\textrm{DRAGON}}$ & $\epsilon_{\textrm{SRIM}}$  \\ \hline
464.2(3) & 40.7(15) & 38.2   \\
442.6(2) & 39.7(15) & 38.8      \\ \hline
441.9(2) & 39.5(15) & 38.8   \\
351.8(3) & 41.5(18) & 39.6     \\
\hline \hline
\end{tabular}
\label{tab:4}
\end{table}

\subsection{GEANT simulations of DRAGON}
\label{sec:geant}

As we have already pointed out in the above, detailed simulations using {\scshape Geant}
are needed to determine the recoil transmission $\mathrm{\eta_{separator}}$ and the
efficiency of the BGO array $\mathrm{\eta_{BGO}}$, which are used to
calculate the reaction yield, and subsequently the resonance strength $\omega \gamma$
as part of the recoil detection efficiency of DRAGON, $\mathrm{\eta_{DRAGON}}$.

The DRAGON {\scshape Geant} simulation toolkit\footnote{The {\scshape Geant} simulation
package of DRAGON can be found at
\href{https://github.com/DRAGON-Collaboration/G3_DRAGON}{https://github.com/DRAGON-Collaboration/G3\_DRAGON}.} has been extensively used for experimental planning, such as in
the study of the $\isotope[12][]{C}(\alpha,\gamma)\isotope[16][]{O}$
reaction~\cite{matei2006measurement}, and its results show
agreement with experimental data to within
1--10\%~\cite{gigliotti2004efficiency}.

The simulation input file includes all the information {\scshape Geant} requires
to perform the simulation such as the energy, spin, lifetime, and
$\gamma$ branching ratios for each nuclear level and in addition the energy and width of
the resonance of interest (see Table~\ref{tab:5} for an overview). For the study
of the \beag~reaction, the nuclear information was adopted from the A=11 evaluation
of~\citet{kelley2012energy}. Specifically for the 1110~keV resonance, since there are no
experimentally measured $\gamma$ branching ratios, we adopted those of the mirror
state in $\isotope[11][]{B}$. For the $\gamma$ ray angular distribution $W(\theta)$, which
affects both the transmission of the recoils and the BGO array
efficiency~\cite{ruiz2014recoil, psaltis6li}, we
calculated all the possible $W(\theta)$ for each cascade, following
the prescription of ~\citet{rose1967angular}. In addition, we changed the gas target pressure in the simulation,
in order to obtain the same stopping power as in the experiment (see Section~\ref{sec:dedx}).

\begin{table*}[hbpt!]
\centering
\caption{Settings of the {\scshape Geant3} simulation for the $\isotope[7][]{Be}(\alpha,\gamma)\isotope[11][]{C}$ data
analysis. Nuclear properties were adopted from~\citet{kelley2012energy}. For the 1110~keV resonance, the branching ratios of the mirror nucleus $\isotope[11][]{B}$ are used.}
\begin{tabular}{llll}
\hline \hline
Quantity & $E_r= 876$~keV & $E_r= 1110$~keV &  $E_r= 1155$~keV\\
\hline
 Excited state lifetime & 0.030 fs & $1.31 \times 10^{-19}$~s & $4.3 \times 10^{-20}$~s \\
 Resonance energy (keV) & 874-878  & 1106--1114~keV & 1153--1157~keV \\
 Particle ($\alpha$) partial width & 12.6~eV & $5$~keV & $15$~keV\\
 $\gamma$ branching ratios & \begin{tabular}{lll} \hline
 $E_x^i$ (MeV) & $E_x^f$ (MeV) & B.R. \\ \hline
 8.420 & 0	     & 93	            \\
     ~ & 4.319	 & 7	 \\
      &          &       \\
      &          &       \\

  \end{tabular} & \begin{tabular}{lll} \hline
 $E_x^i$ (MeV) & $E_x^f$ (MeV) & B.R. \\ \hline
 8.654 & 0     & 0.9 $\pm$ 0.3  \\
                & 4.319     & 86.6 $\pm$ 2.3 \\
                & 6.478     & 12.5 $\pm$ 1.1 \\
                 &          &       \\
  \end{tabular}    &
  \begin{tabular}{lll} \hline
 $E_x^i$ (MeV) & $E_x^f$ (MeV) & B.R. \\ \hline
   8.699 & 0   	     & 42 $\pm$ 10      \\
    	       & 4.319     & 42 $\pm$ 10 	    \\
               & 4.804     & 2.4 $\pm$ 1.5    \\
               & 6.478     & 13.6 $\pm$ 4.6   \\
  \end{tabular} \\
\hline \hline
\end{tabular}
\label{tab:5}
\end{table*}

We performed simulations for each resonance energy within its uncertainty,
$\pm$2, $\pm$4 and $\pm 2$~keV for the 1155, 1110 and 876~keV resonance respectively.
The final results used in the data analysis are the averages
of these simulations and the systematic uncertainty is attributed mainly to
the uncertainty in the $\gamma$ branching ratios and the range of possible $\gamma$ angular distributions.

\begin{figure}[ht!]
    \centering
    \includegraphics[width=.45\textwidth]{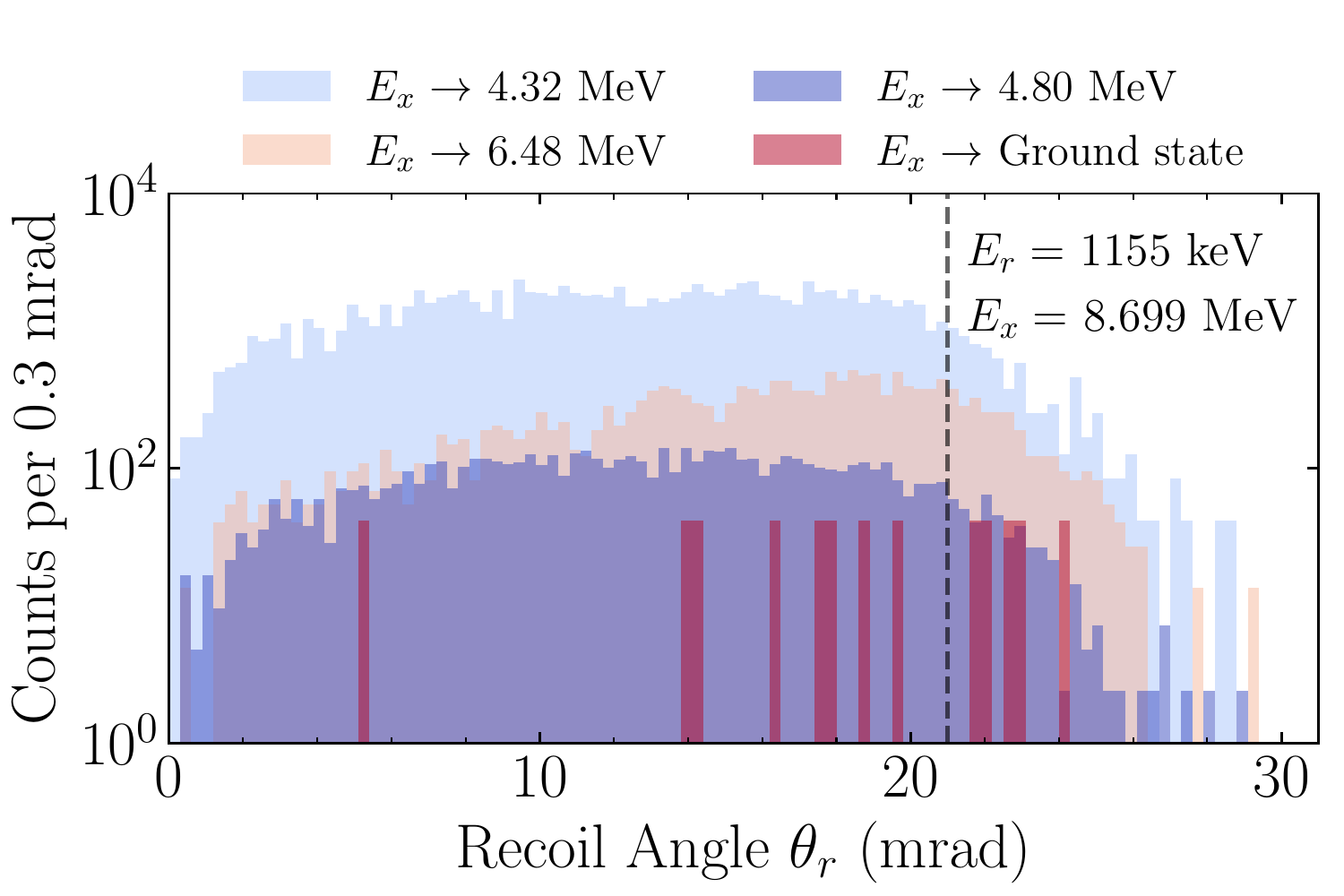}
    \caption{Angular distribution for recoils of the $E_r$= 1155~keV resonance that hit the focal plane detector (DSSSD) using {\scshape Geant}. The contributions from different cascades are shown. The vertical dashed line defines the angular acceptance of DRAGON, $\theta_{DRAGON} = 21$~mrad.}
\label{fig:7}
\end{figure}

Figure~\ref{fig:7} shows the results from a
simulation of the $E_r$= 1155~keV resonance. It is evident that transitions
with a cascade of multiple $\gamma$ rays, such as the 8.699~MeV $\rightarrow$ 4.32~MeV
provide more favourable conditions for transmission through the separator,
since their recoil angular distribution from multiple decay vectors averages out with
resulting maximum intensity at lower angles, as~\citet{ruiz2014recoil} argue.

\section{Results}
\label{sec:5}

In the following, we shall discuss the results
from each resonance studied in the present work and then
present the uncertainties and the calculation of the new
thermonuclear reaction rate $N_A \langle \sigma v \rangle$.

\subsection{Strength of the 1155~keV resonance}

For the highest energy resonance we studied in the present work,
1155~keV, we detected a strong, background--free signal,
as is evident in the PID plot of Figure~\ref{fig:4}.
Clusters of 33 and 49 recoil events in coincidence and singles
modes were detected, despite the low recoil transmission through the separator ($\mathrm{\eta_{separator}}$ = 0.141(28), see also
Table~\ref{tab:8}).

\begin{figure}[hbpt!]
    \centering
    \includegraphics[width=.5\textwidth]{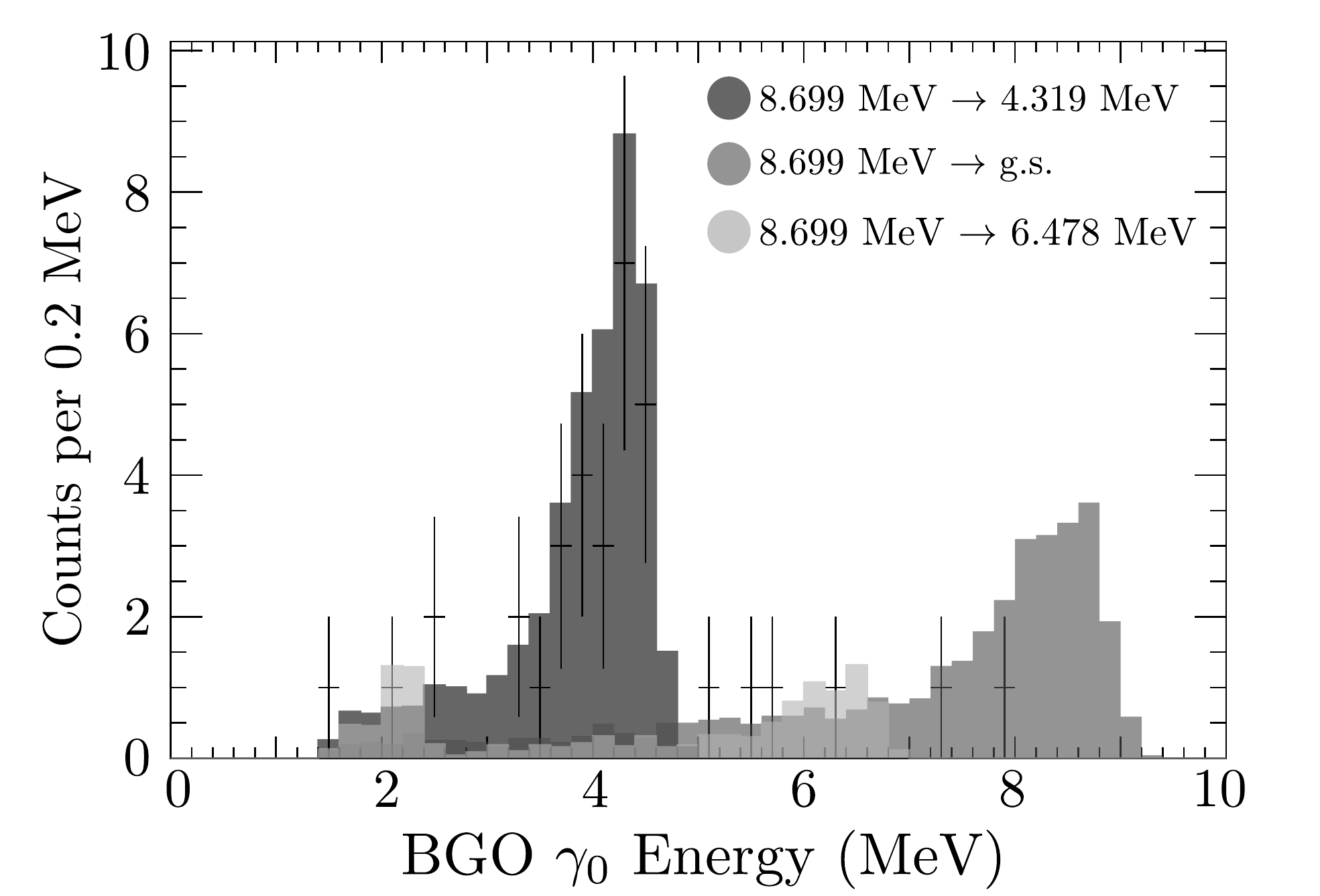}
    \caption{Comparison between the experimental data (points) and the {\scshape Geant}
    simulation (histograms) for a BGO $\gamma_0$ ray energy spectrum from the 1155~keV resonance. The two dominant $\gamma$ transitions to the
    $E_x= 4.319$~MeV and the ground state can be clearly seen in the simulation results. We do not depict the weak
    transition to the $E_x= 4.804$~MeV state. See the text for details.
    }
\label{fig:8}
\end{figure}

The {\scshape Geant} simulations we performed for this resonance are in very
good agreement with the experimental results, as Figure~\ref{fig:8} shows (see also Figure 2 of Ref.~\cite{psaltis_prl} for the same resonance).
It is worth noting that DRAGON is more sensitive to the detection of recoils
that $\gamma$ decay to the $E_x= 4.319$~MeV state, compared to the ground
state, as we discussed in Section~\ref{sec:geant}.

The final result for the resonance strength is
$\omega \gamma_{1155}= 1.73 \pm 0.25(stat.) \pm 0.40(syst.)$~eV was adopted from the singles analysis
and had a smaller uncertainty compared to the coincidence analysis result.
However, the two results are in agreement (see also Table~\ref{tab:6}).

\subsection{Strength of the 1110~keV resonance}
\label{sec:er1110}

The 1110~keV resonance was studied in two independent experimental runs, due
to the low recoil yields.
We detected $14^{+4.3}_{-3.7}$ and $9^{+3.8}_{-2.7}$ events in singles,
with different integrated beam fluxes $1.76(5) \times 10^{13}$ and
$1.53(5) \times 10^{13}$ ions, respectively (see also Table~\ref{tab:3}).
The asymmetric uncertainty in the amount of detected recoils was calculated according
to the prescription of~\citet{feldman1998unified} for a poissonian signal with
zero background, as we can see from the PID plots of Figure~\ref{fig:4}.

\begin{figure}[hbpt!]
    \centering
    \includegraphics[width=.45\textwidth]{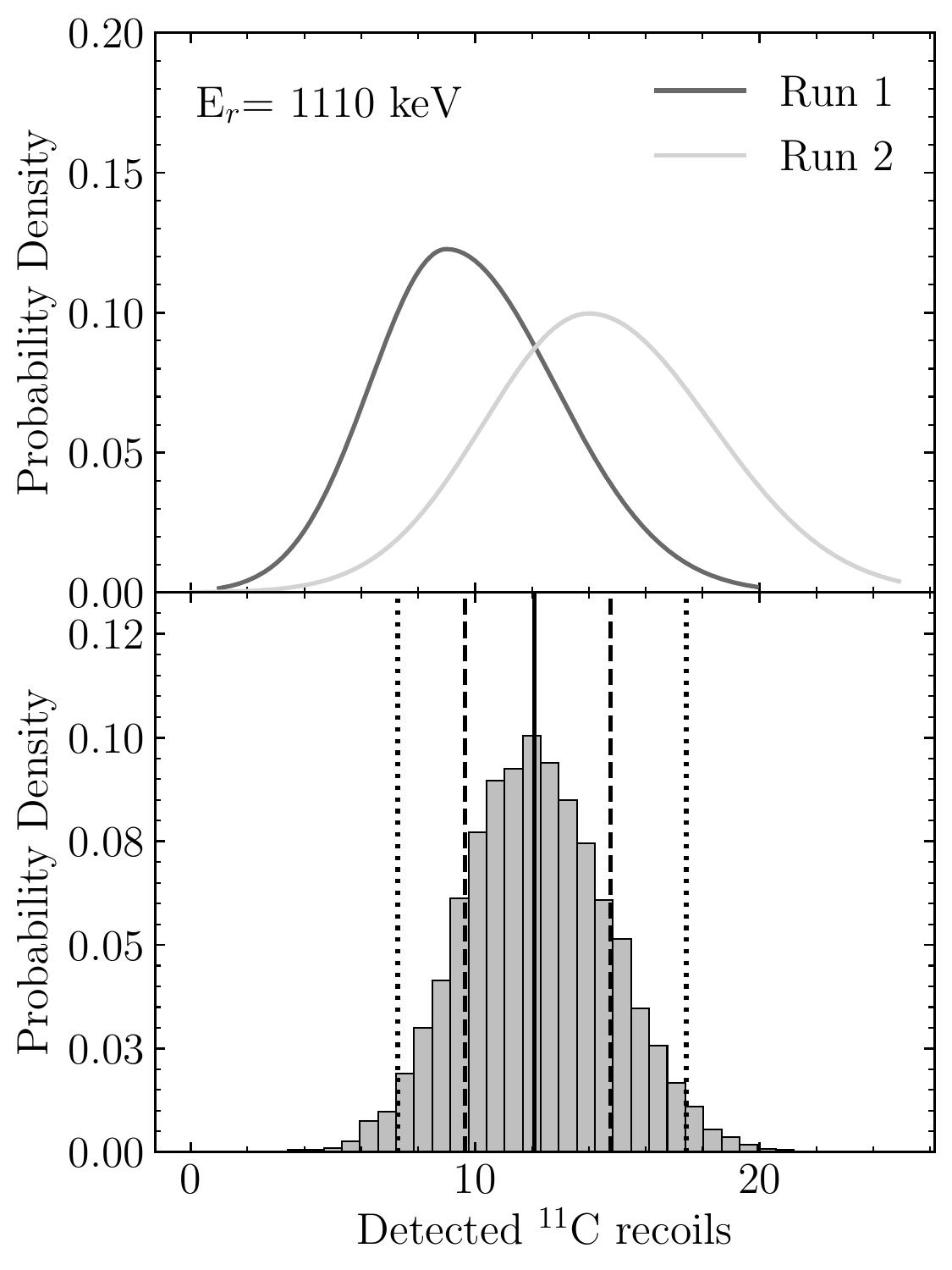}
    \caption{(Top) Individual probability distributions for the detected $\isotope[11][]{C}$ recoils in singles mode from the two independent measurements of the  1110~keV resonance. (Bottom) Combined probability distribution from the same measurements. The solid line shows the central value, while the dashed and dotted lines show the 1 and $2\sigma$ uncertainties, respectively. See the text for details.}
\label{fig:9}
\end{figure}

To account for the asymmetric uncertainties and provide a realistic
statistical uncertainty for the number of detected $\isotope[11][]{C}$ recoils, we
proceeded as follows: we first used Fechner's two--piece normal distribution~\citep{wallis2014two} for the two
independent runs, using the~\citet{feldman1998unified} prescription for the variances (see
Figure~\ref{fig:9} - Top). After that, we created a combined probability distribution by calculating averages by randomly sampling from the two individual distributions. The final results for the detected $\isotope[11][]{C}$ recoils and their respected 1 and $2\sigma$ uncertainties are then calculated from the combined distribution. We find $12.1^{+2.7}_{-2.5}$ ($1\sigma$) and $^{+5.3}_{-4.8}$ ($2\sigma$) events for the 1110~keV resonance in singles, corresponding to $^{+22.3}_{-20.7}\%$ ($1\sigma$) and $^{+43.8}_{-39.7}\%$($2\sigma$) statistical uncertainty, respectively as we show in in Figure~\ref{fig:9}.

The resonance strengths resulting from singles and coincidence analysis are
$\mathrm{\omega \gamma_{1110, singles}} = 125 ^{+27}_{-25}(stat.) \pm 15(syst.)$~meV and $\mathrm{\omega \gamma_{1110, coinc}}= 161 ^{+43}_{-41}(stat.) \pm 24(syst.)$~meV, respectively, from which we choose the former
as the final result. The large difference compared to the 1155~keV resonance strength -- almost an order of magnitude -- can be attributed to
the difference in the orbital angular momentum, $\ell_\alpha = 1$ for the former state and $\ell_\alpha =2$ for the latter (see Table~\ref{tab:1}).

\subsection{Strength of the 876~keV resonance}

For the lowest energy in this study, the maximum recoil angle is
$\theta_{r,max}$= 47~mrad, which is the largest ever attempted by DRAGON\footnote{The previous largest maximum recoil angle was $\theta_{r,max}$= 33~mrad in the study of
the $\isotope[12][]{C}(\isotope[16][]{O},\gamma)\isotope[28][]{Si}$ reaction~\cite{PhysRevC.85.034333}.}.
Nevertheless, one can see a clear signal in the PID plots (see Figures~\ref{fig:4} and \ref{fig:5}). We detected 13$^{+4.3}_{-3.7}$ $\isotope[11][]{C}$ recoil events
corresponding to $^{+33.1}_{-28.5} \%$ 1$\sigma$ statistical uncertainty, following the~\citet{feldman1998unified} prescription for a poissonian signal with zero background.

Our final result for its strength from the singles analysis is
$\omega \gamma_{876}= 3.00^{+0.81}_{-0.72} (stat.) \pm 0.61(syst.)$~eV. We calculated the weighted average of our measurement and the value by~\citet{hardie1984resonant} to get the adopted resonance strength
$\omega \gamma_{876}= 3.61(50)$~eV, which will be used for the calculation of the thermonuclear reaction
rate in Section~\ref{sec:rate}. It is worth noting that for all resonances the results for the
strength $\omega \gamma$ agree both in coincidence and singles analysis modes.

\begin{table*}[ht!]
\centering
\caption{
Resonance strengths $\omega\gamma$ of the \beag~reaction resonances from the literature~\cite{hardie1984resonant, kelley2012energy}
and the present work (in singles and coincidences modes) that were  used for the calculation of the new thermonuclear reaction rate. All
results are reported in eV and the statistical and systematic uncertainties are presented separately. The adopted value for the 876~keV resonance strength is the weighted average of our singles measurement and the one from~\citet{hardie1984resonant}. See the text for details.}
\begin{tabular}{ccccc}
\hline \hline
$\mathrm{E_r}$~(keV) & Literature & Singles & Coincidences & Adopted \\ \hline
561 & $0.331(41)$ & $\cdots$ & $\cdots$ & $0.331(41)$ \\
876 & $3.80(57)$ & $3.00^{+0.81}_{-0.72}(stat.)\pm 0.61(syst.)$ & $3.91^{+1.29}_{-1.10}(stat.)\pm 1.18(syst.)$ & $3.61(50)$\\
1110 & $\cdots$ &  $0.125 ^{+0.027}_{-0.025}(stat.) \pm 0.015(syst.)$ &  $0.161 ^{+0.043}_{-0.041}(stat.) \pm 0.024(syst.)$ &  $0.125(31)$\\
1155 & $\cdots$ &   $1.73 \pm 0.25(stat.) \pm 0.40(syst.)$ & $1.79 \pm 0.33(stat.) \pm 0.42(syst.)$ & $1.73(47)$\\
\hline \hline
\end{tabular}
\label{tab:6}
\end{table*}

Also, from a nuclear structure standpoint,  our
results for the two previously unknown resonance strengths are in very good agreement with
their $\isotope[7][]{Li}(\alpha,\gamma)\isotope[11][]{B}$ analogs~\citep{kelley2012energy}, namely $E_x= 9.182$ ($7/2^+$) and 9.271~MeV ($5/2^+$), as they are
shown in Table~\ref{tab:7}. The $5/2^-$ state $E_x= 8.921$ in $\isotope[11][]{B}$ is the only exception, were its resonance strength
differs more than two orders of magnitude to its $\isotope[11][]{C}$ analog.

\begin{table}[ht!]
\centering
\caption[Comparison of $\omega \gamma$ for $^7$Be($\alpha,\gamma$)$^{11}$C and $^7$Li($\alpha,\gamma$)$^{11}$B]{Comparison of resonance strengths $\omega \gamma$ for analog states in $^7$Be($\alpha,\gamma$)$^{11}$C and $^7$Li($\alpha,\gamma$)$^{11}$B reactions. Literature data were
taken from~\citet{kelley2012energy}.}
\begin{tabular}{cccc}
\hline \hline
$J^\pi$ & Nucleus  & E\textsubscript{x}(MeV) & $\omega \gamma$ (eV)\\ \hline
5/2$^-$ & $\isotope[11][]{B}$ & 8.921(1) & $(8.8 \pm 1.4) \times 10^{-3}$\\
5/2$^-$ & $\isotope[11][]{C}$ & 8.420(2) & $3.61(50)$\\\hline
7/2$^+$ & $\isotope[11][]{B}$ & 9.182(2) & $0.303(26)$\\
7/2$^+$ & $\isotope[11][]{C}$ & 8.654(4) & $0.125(31)$\\\hline
5/2$^+$ & $\isotope[11][]{B}$ & 9.271(2) & $1.72(24)$\\
5/2$^+$ & $\isotope[11][]{C}$ & 8.699(2) & $1.73(47)$\\
\hline \hline
\end{tabular}
\label{tab:7}
\end{table}

\subsection{Uncertainties}
\label{sec:uncert}

The uncertainties of the final results of the resonance strengths of this
study are of systematic and statistical nature. The former are dominated
by the efficiencies of the BGO array ($\eta_{BGO}$) and the recoil
transmission through the separator ($\eta_{separator}$),
which are inferred from {\scshape Geant} simulations
(Section~\ref{sec:geant})~\cite{ruiz2014recoil}. Other sources of systematic
uncertainty are the MCP detection efficiency, the stopping power
measurements and the charge state fractions.

Furthermore, the statistical
uncertainties are due to the low recoil detection yield, caused by
the very low transmission of the recoils through the separator, but this parameter is well understood and quantified. As we already pointed out,
for the 1110 and 876~keV resonances, we used the prescription
of~\citet{feldman1998unified} for poissonian signals in zero background to extract the statistical uncertainties.

Table~\ref{tab:8} shows a detailed breakdown of the
uncertainties for each of the three resonances we measured in the present
study. Note that for the 1110~keV resonance the uncertainty of the
average final result was calculated using the procedure discussed in Section~\ref{sec:er1110}.

%compare BGO uncertainty simulation vs real?
\begin{table*}[hbpt!]
\centering
\caption{Values given with uncertainties for the quantities used to calculate the resonance strengths in singles and coincidences modes. The relative errors are quoted in parentheses.}
\begin{tabular}{c|c|c|c|c}
\hline \hline
Source & 1155~keV & 1110~keV (Run 1) & 1110~keV (Run 2) & 876~keV \\ \hline
Detected recoils, $\mathrm{N_{rec}^{singles}}$ & 49(7) (14.3\%) & $14^{+4.3}_{-3.7}$ ($^{+31}_{-26}$\%) & $9^{+3.8}_{-2.7}$ ($^{+42}_{-30}$\%) & $16^{+4.3}_{-3.8}$ ($^{+27}_{-24}$\%)\\
Detected recoils, $\mathrm{N_{rec}^{coinc}}$ & 33(6) (18.2\%)& $9^{+3.8}_{-2.7}$ ($^{+42}_{-30}$\%) & $7^{+3.3}_{-2.8}$ ($^{+47}_{-40}$\%) & $13^{+4.3}_{-3.7}$ ($^{+33}_{-28}$\%) \\
Charge state fraction, $f_q$ & 0.40(1) (2.5\%) & 0.41(1) (2.4\%) & 0.41(1) (2.4\%) & 0.41(1) (2.4\%) \\
Beam particles, $\mathrm{N_{beam} \times 10^{13}}$ & 1.07(2) (1.9\%) & 1.76(5) (2.8\%) & 1.53(5) (3.2\%)& 2.12(3) (1.41\%) \\
BGO efficiency, $\mathrm{\eta_{BGO}}$ & 0.77(1) (1.3\%) & 0.81(7) (8.6\%) & 0.81(7) (8.6\%) & 0.80(18)(22.5\%) \\
Separator transmission, $\mathrm{\eta_{separator}}$ & 0.141(28) (19.9\%)  & 0.266(18) (6.8\%)  & 0.266(18) (6.8\%) & 0.016(3) (18.8\%)\\
MCP efficiency\footnote{The MCP efficiency includes both the detection efficiency of the system and also the transmission of the recoils through the thin carbon foil that creates the secondary electrons that the MCP detects.}, $\mathrm{\eta_{MCP}}$ & 0.545(59) (10.8\%) & 0.650(61) (9.4\%) &  0.321(25) (7.8\%)& 0.351(19) (5.4\%)\\
Live time, $\mathrm{\eta_{live}^{singles}}$ & 0.95409(5) (0.005\%) & 0.95777(5) (0.005\%) & 0.99099(5) (0.005\%) & 0.93408(5) (0.005\%)\\
Live time, $\mathrm{\eta_{live}^{coinc}}$ & 0.80381(4) (0.005\%) & 0.80434(4) (0.005\%) &
0.82156(4) (0.005\%) & 0.81571 (0.005\%) \\
Stopping power, $\epsilon$ (eV/($10^{15}$/cm$^2$)) & 40.7(15) (3.7\%) & 39.7(15) (3.8\%) & 39.5(15) (3.8\%) & 41.5(18) (4.3\%)\\
Beam energy (A keV) & 462.2(3) (0.06\%) &442.6(2) (0.05\%)  & 441.8(2) (0.05\%) & 351.8(3) (0.09\%) \\ \hline
\textbf{Total uncertainty in singles} &  14.3\% (stat.) &  $^{+31}_{-26}\%$ (stat.) & $^{+42}_{-30}\%$ (stat.) & $^{+27}_{-24}\%$ (stat.) \\
(statistical \& systematic)  & 23.2\% (syst.)& 12.8\% (syst.) & 11.8\% (syst.)& 20.2\% (syst.) \\ \hline
\textbf{Total uncertainty in coincidences}& 18.2\% (stat.) & $^{+42}_{-30}$\% (stat.)& $^{+47}_{-40}$\% (stat.) & $^{+33}_{-28}$\% (stat.)\\
(statistical \& systematic) &  23.2\% (syst.)& 15.4\% (syst.) & 14.6\% (syst.)& 30.3\% (syst.) \\
\hline \hline
\end{tabular}
\label{tab:8}
\end{table*}

\subsection{Thermonuclear Reaction Rate}
\label{sec:rate}

The new \beag~ thermonuclear reaction rate was calculated using the RatesMC\footnote{The RatesMC code to calculate thermonuclear reaction rates can be found at \href{https://github.com/rlongland/RatesMC}{https://github.com/rlongland/RatesMC}.} code~\citep{longland2010charged}.
Within the RatesMC framework, each nuclear
physics input quantity (\emph{e.g.} resonance energy and resonance strength)
has an assigned probability density function (PDF). The code samples these functions
randomly many times ($> 10^3$) using a Monte Carlo algorithm and
outputs reaction rates and associated rate probability densities.
According to the central limit theorem, a random variable that is determined by the product
of many factors will be distributed according to a lognormal density
function~\cite{gaddum1945lognormal, longland2010charged, iliadis2015statistical}. Using a lognormal PDF,
the ``low'', ``recommended'', and ``high'' Monte
Carlo rates are the 16\textsuperscript{th}, 50\textsuperscript{th} (median), and
84\textsuperscript{th} percentile respectively of the cumulative reaction rate
distribution. In Table~\ref{tab:9} we present the adopted thermonuclear reaction rate for \beag.

\begin{table}[hbpt!]
\caption[Total thermonuclear reaction rates for $^7$Be($\alpha,\gamma$)$^{11}$C. The rate below T$ < 0.012$~GK is zero. The rates are expressed in units of cm$^3$ mol$^{-1}$ s$^{-1}$.]{Total thermonuclear reaction rates for $^7$Be($\alpha,\gamma$)$^{11}$C. The rate below T$ < 0.012$~GK is zero. The rates are expressed in units of cm$^3$ mol$^{-1}$ s$^{-1}$. Columns 2,3 and 4 list the 16th, 50th and 86th percentiles of the total rate probability density (PDF) at given temperatures. \textit{``f.u''} is the factor uncertainty, and is obtained from the 16th and 84th percentiles.
}
\begin{ruledtabular}
\begin{tabular}{ccccc}
T (GK)& Low  & Median  & High  &   f.u.\\  \hline
0.012 & $3.463 \times 10^{-34}$ & $8.835 \times 10^{-34}$ &
      $2.348 \times 10^{-33}$ & 2.590 \\
0.013 & $4.535 \times 10^{-33}$ & $1.153 \times 10^{-32}$ &
      $3.066 \times 10^{-32}$ & 2.588 \\
0.014 & $4.594  \times 10^{-32}$ & $1.168  \times 10^{-31}$ &
      $3.097 \times 10^{-31}$ & 2.586 \\
0.015 & $3.758 \times 10^{-31}$ & $9.552 \times 10^{-31}$ &
      $2.528 \times 10^{-30}$ & 2.584 \\
0.016 & $2.563 \times 10^{-30}$ & $6.511 \times 10^{-30}$ &
      $1.725 \times 10^{-29}$ & 2.583 \\
0.018 & $7.618 \times 10^{-29}$ & $1.935 \times 10^{-28}$&
      $5.129 \times 10^{-28}$ & 2.579 \\
0.020 & $1.407 \times 10^{-27}$ & $3.572 \times 10^{-27}$ &
      $9.463 \times 10^{-27}$ & 2.576 \\
0.025 & $4.793 \times 10^{-25}$ & $1.208 \times 10^{-24}$ &
      $3.195 \times 10^{-24}$ & 2.568 \\
0.030 & $4.012 \times 10^{-23}$ & $1.008 \times 10^{-22}$ &
     $2.665 \times 10^{-22}$ & 2.560 \\
0.040 & $2.466 \times 10^{-20}$ & $6.147 \times 10^{-20}$ &
      $1.616 \times 10^{-19}$ & 2.544 \\
0.050 & $2.309 \times 10^{-18}$ & $5.693 \times 10^{-18}$ &
      $1.495 \times 10^{-17}$ & 2.528 \\
0.060 & $7.225 \times 10^{-17}$ & $1.766 \times 10^{-16}$ &
      $4.628 \times 10^{-16}$ & 2.513 \\
0.070 & $1.117 \times 10^{-15}$ & $2.705 \times 10^{-15}$ &
      $7.081 \times 10^{-15}$ & 2.498 \\
0.080 & $1.057 \times 10^{-14}$ & $2.548 \times 10^{-14}$ &
      $6.647 \times 10^{-14}$ & 2.483 \\
0.090 & $6.993 \times 10^{-14}$ & $1.681 \times 10^{-13}$ &
      $4.372 \times 10^{-13}$ & 2.468 \\
0.100 & $3.570 \times 10^{-13}$ & $8.501 \times 10^{-13}$ &
      $2.201 \times 10^{-12}$ & 2.453 \\
0.110 & $1.472 \times 10^{-12}$ & $3.490 \times 10^{-12}$ &
      $8.984 \times 10^{-12}$ & 2.439 \\
0.120 & $5.172 \times 10^{-12}$ & $1.216 \times 10^{-11}$ &
      $3.106 \times 10^{-11}$ & 2.424 \\
0.130 & $1.580 \times 10^{-11}$ & $3.679 \times 10^{-11}$ &
      $9.412  \times 10^{-11}$ & 2.410 \\
0.140 & $4.322 \times 10^{-11}$ & $9.999 \times 10^{-11}$ &
      $2.548 \times 10^{-10}$ & 2.396 \\
0.150 & $1.078 \times 10^{-10}$ & $2.473 \times 10^{-10}$ &
      $6.275 \times 10^{-10}$ & 2.382 \\
0.160 & $2.482  \times 10^{-10}$ & $5.656 \times 10^{-10}$ &
      $1.429 \times 10^{-9}$ & 2.367 \\
0.180 & $1.116 \times 10^{-9}$ & $2.459 \times 10^{-9}$ &
      $6.115 \times 10^{-9}$ & 2.310 \\
0.200 & $4.918 \times 10^{-9}$ & $9.604 \times 10^{-9}$ &
      $2.208 \times 10^{-8}$ & 2.096 \\
0.250 & $5.348 \times 10^{-7}$ & $6.305 \times 10^{-7}$ &
      $7.894 \times 10^{-7}$ & 1.251 \\
0.300 & $2.666 \times 10^{-5}$ & $3.051 \times 10^{-5}$ &
      $3.506 \times 10^{-5}$ & 1.147 \\
0.350 & $4.586 \times 10^{-4}$ & $5.215 \times 10^{-4}$ &
      $5.955 \times 10^{-4}$ & 1.142 \\
0.400 & $3.813 \times 10^{-3}$ & $4.334 \times 10^{-3}$ &
      $4.923 \times 10^{-3}$ & 1.139 \\
0.450 & $1.950 \times 10^{-2}$ & $2.209 \times 10^{-2}$ &
      $2.507 \times 10^{-2}$ & 1.137 \\
0.500 & $7.106 \times 10^{-2}$ & $8.022 \times 10^{-2}$ &
      $9.077 \times 10^{-2}$ & 1.135 \\
0.600 & $4.816 \times 10^{-1}$ & $5.413 \times 10^{-1}$ &
      $6.105 \times 10^{-1}$ & 1.130 \\
0.700 & $1.860 \times 10^0$ & $2.083 \times 10^0$ &
      $2.340 \times 10^0$ & 1.125 \\
0.800 & $5.136  \times 10^0$ & $5.726 \times 10^0$ &
      $6.390 \times 10^0$ & 1.118 \\
0.900 & $1.139  \times 10^1$ & $1.265  \times 10^1$ &
     $1.401  \times 10^1$ & 1.111 \\
1.000 & $2.178  \times 10^1$ & $2.407  \times 10^1$ &
      $2.651  \times 10^1$ & 1.105 \\
1.250 & $7.217 \times 10^1$ & $7.907 \times 10^1$ &
      $8.638  \times 10^1$ & 1.095 \\
1.500 & $1.633 \times 10^2$ & $1.792 \times 10^2$ &
      $1.953 \times 10^2$ & 1.094 \\
1.750 & $2.946 \times 10^2$ & $3.236 \times 10^2$ &
      $3.538 \times 10^2$ & 1.095 \\
2.000 & $4.578 \times 10^2$ & $5.041 \times 10^2$ &
      $5.522 \times 10^2$ & 1.098 \\
2.500 & $8.379\times 10^2$ & $9.262 \times 10^2$ &
      $1.021 \times 10^3$ & 1.103 \\
3.000 & $1.235 \times 10^3$ & $1.370 \times 10^3$ &
      $1.516 \times 10^3$ & 1.107 \\
3.500 & $1.612 \times 10^3$ & $1.801 \times 10^3$ &
      $2.003 \times 10^3$ & 1.116 \\
4.000 & $1.955 \times 10^3$ & $2.195 \times 10^3$ &
      $2.468 \times 10^3$ & 1.129 \\
5.000 & $2.515 \times 10^3$ & $2.876 \times 10^3$ &
      $3.332 \times 10^3$ & 1.167 \\
6.000 & $2.924 \times 10^3$ & $3.398 \times 10^3$ &
      $4.121 \times 10^3$ & 1.210 \\
7.000 & $3.202 \times 10^3$ & $3.812 \times 10^3$ &
      $4.777 \times 10^3$ & 1.250 \\
8.000 & $3.386 \times 10^3$ & $4.112 \times 10^3$ &
      $5.325 \times 10^3$ & 1.284 \\
9.000 & $3.503 \times 10^3$ & $4.324 \times 10^3$ &
      $5.765 \times 10^3$ & 1.312 \\
10.000 & $3.558 \times 10^3$ & $4.460 \times 10^3$ &
      $6.079 \times 10^3$ & 1.335 \\

\end{tabular}
\end{ruledtabular}
   \label{tab:9}
\end{table}

%Discussion about the calculation of the new rate (RatesMC)
For our calculation, we used resonance parameters as reported in~\citet{kelley2012energy}. More specifically, we
included the contribution of the sub--threshold resonance at $\mathrm{E_x}$ = 7.4997~MeV ($E_r= -43.9$~keV) using a $\gamma$ partial width from the mirror state in $\isotope[11][]{B}$ ($\Gamma_\gamma = 1.14(4)$~eV), and assumed a reduced $\alpha$ width of 1. According to~\citet{descouvemont19957be} this resonance can dominate the reaction rate for T$<$ 0.3~GK , which can affect the evolution of Population III stars via the hot \textit{pp}--chains
~\cite{wiescher1989hot} and the production of $\isotope[7][]{Li}$ in classical novae~\cite{hernanz1996synthesis}.
In addition to the narrow resonances at 561, 876, 1110~\&~1155~keV, we also included contributions from the broad resonances at 2101, 2236, 2426~\&~2539~keV (see Table~\ref{tab:1} for details). For the $\gamma$ partial widths of the latter states, since we used values from the mirror analog $\isotope[11][]{B}$, we assigned them a factor of 2 uncertainty.

\begin{figure}
    \centering
    \includegraphics[width=.5\textwidth]{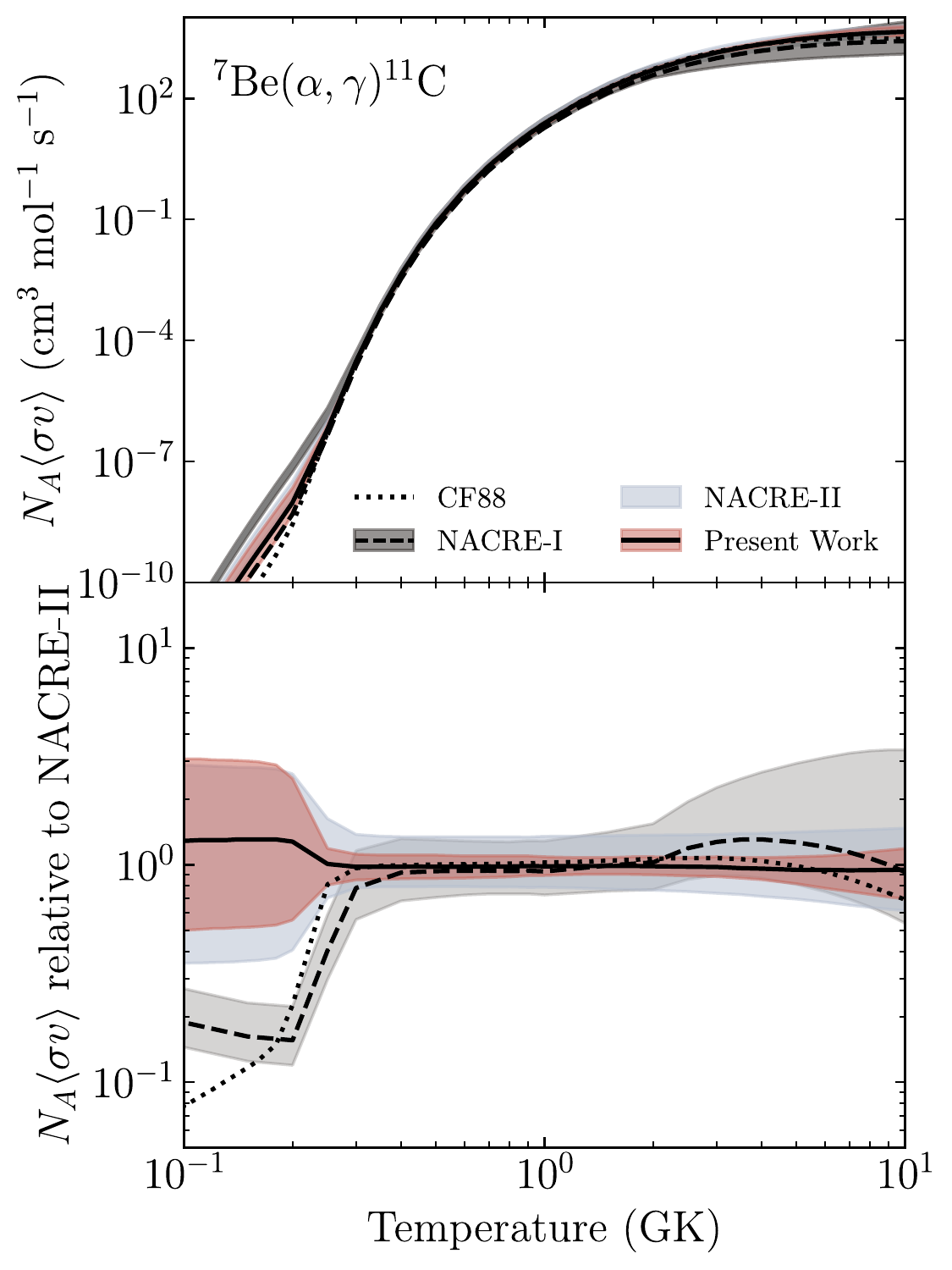}
    \caption{The new $\isotope[7][]{Be}(\alpha,\gamma)\isotope[11][]{C}$ reaction rate for T= 0.1--10 GK compared to the rates by References~\cite{angulo1999compilation, xu2013nacre,caughlan1988thermonuclear} over the same temperature region.}
\label{fig:10}
\end{figure}

The new reaction rate uncertainty has been decreased to $\sim 9.4-10.7\%$ over T= 1.5--3~GK, the relevant temperature window for $\nu p$--process nucleosynthesis, compared to factors of 1.76-1.91 of the NACRE--II compilation. While our new rate includes the new measurements of the 1110 and 1155~keV resonance strengths and the updated adopted value for $\omega\gamma_{876}$, it is worth noting that this decrease in the thermonuclear reaction rate is mainly caused from using a Monte Carlo error propagation of the relevant quantities (\textit{e.g.} $E_r$, $\omega\gamma$ \textit{etc.})~\citep{longland2010charged}, and not by the individual contribution of the previously unmeasured resonance strengths.
In Figure~\ref{fig:10} we compare the new thermonuclear reaction rate to the NACRE rate~\citep{angulo1999compilation,xu2013nacre} and an older reaction rate compilation from~\citet{caughlan1988thermonuclear} (CF88).

\begin{figure}[ht!]
    \centering
    \includegraphics[width=.5\textwidth]{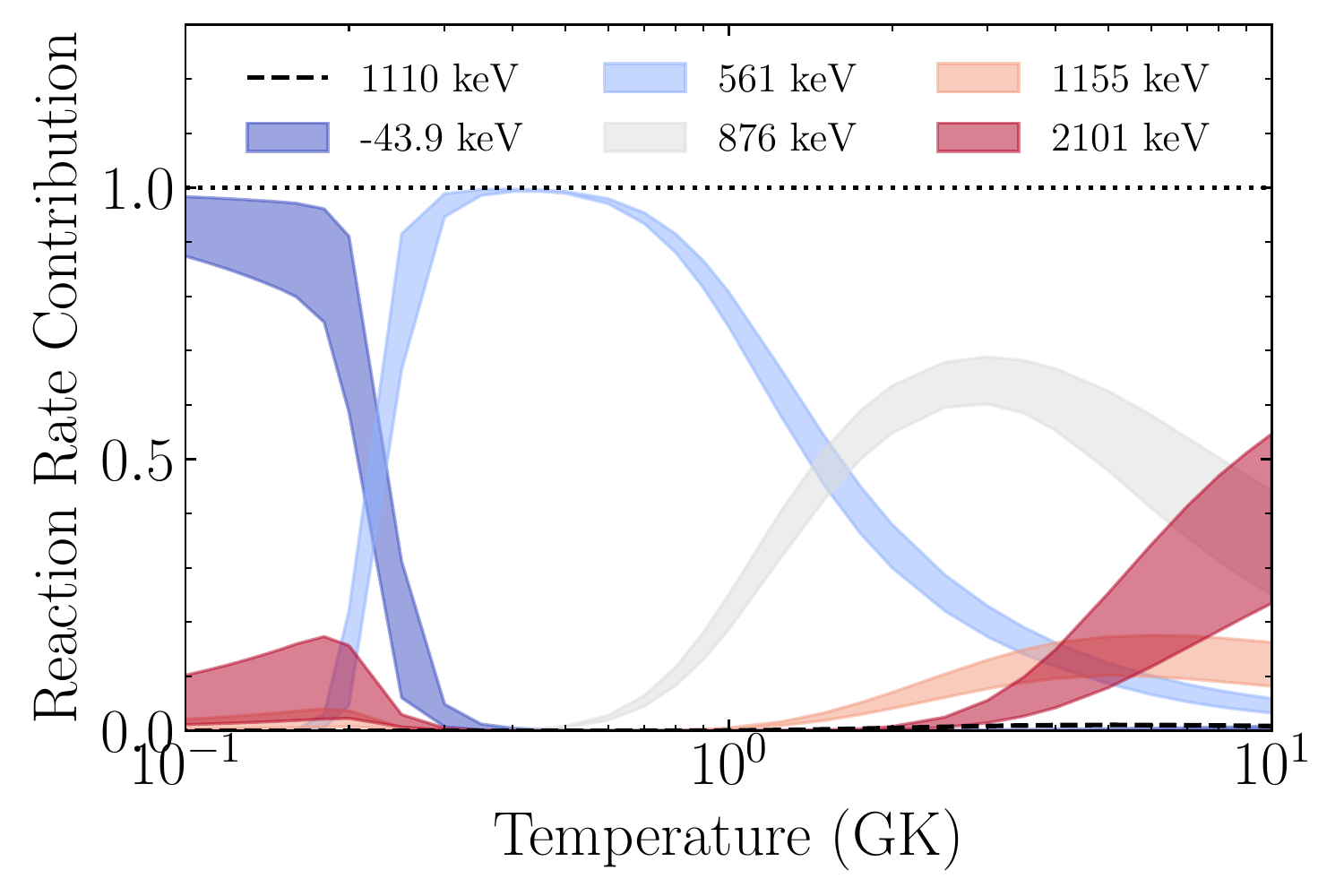}
    \caption{Resonant contributions to the \beag~ thermonuclear reaction rate. The dotted line
    at the bottom right corner shows the contribution of the 1110~keV resonance.
    }
\label{fig:11}
\end{figure}

Figure~\ref{fig:11} shows the individual resonant contributions to the total \beag~reaction rate. For temperatures $T \lessapprox 0.2$~GK, the sub--threshold resonance at -43.9~keV dominates the reaction rate, while for $\mathrm{0.2~GK < T < 1.0~GK}$ the 561~keV contributes the most, since $\Gamma_\alpha \gg \Gamma_\gamma$ (see Table~\ref{tab:1}) and it is the lowest--lying energy resonance~\citep[see the discussion in Ref.][Chapter 3]{iliadis2015nuclear}. For the temperatures relevant for the $\nu p$--process, the 876~keV has a $\sim 60\%$ contribution to the total rate, followed by the 561~keV with $\sim 30\%$. The 1155~keV resonance has a $ \lessapprox 10\%$ contribution, while the 1110~keV contributes negligibly to the total reaction rate.

The effect of the new reaction rate will be studied extensively in a future publication, taking
into account new measurements of the $\isotope[10][]{B}(\alpha,p)\isotope[13][]{C}$
reaction~\cite{liu2020low}, which was also included in the sensitivity study
of~\citet{wanajo2011uncertainties}, and the
$\isotope[59][]{Cu}(p,\alpha)\isotope[56][]{Ni}$~\cite{PhysRevC.104.L042801} which may be responsible for the Ni--Cu
cycle~\cite{arcones2012impact}.

\begin{figure}
    \centering
    \includegraphics[width=.5\textwidth]{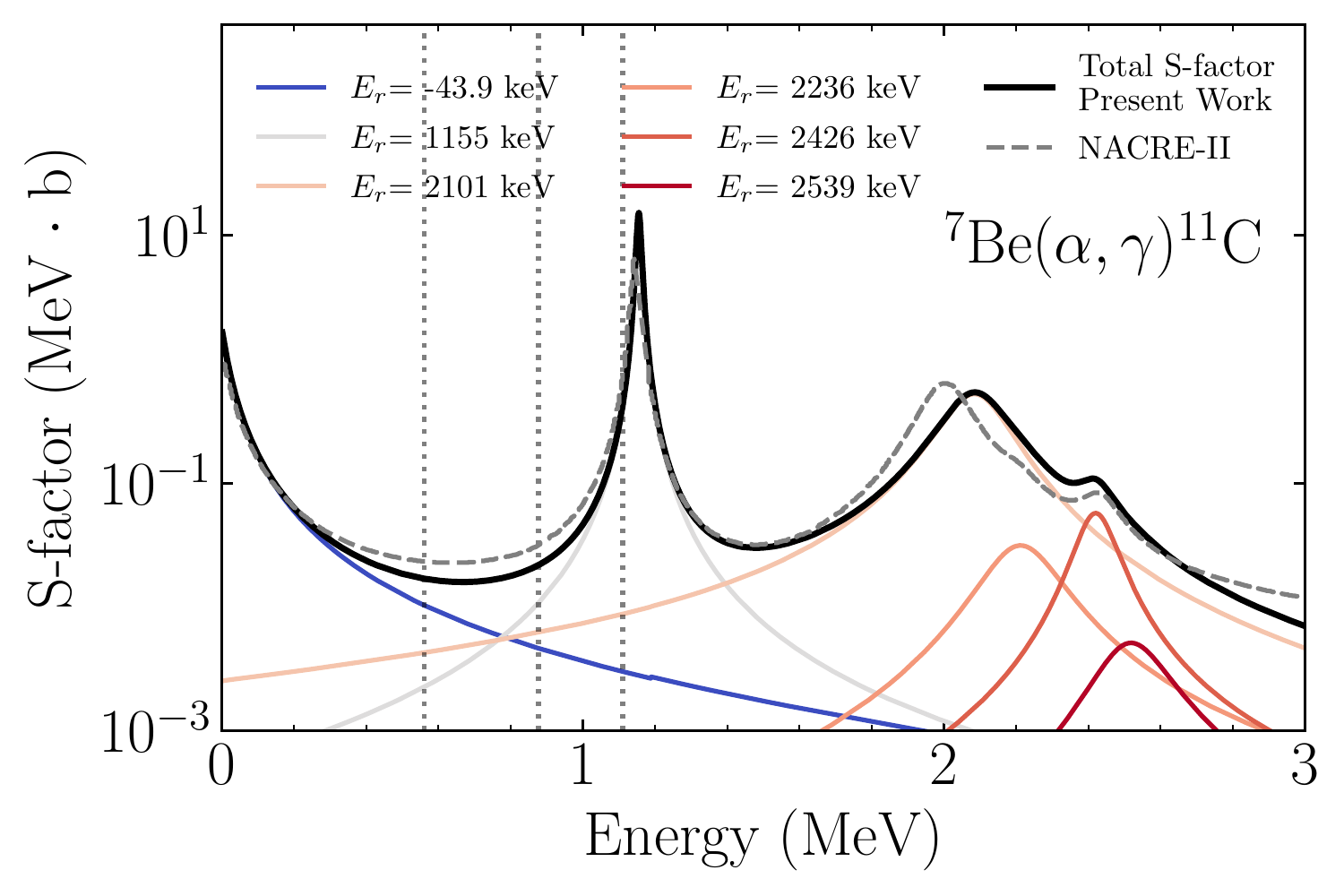}
    \caption{The astrophysical $S$--factor for the $\isotope[7][]{Be}(\alpha,\gamma)\isotope[11][]{C}$ reaction based on our RatesMC calculations. Contributions from different resonances are shown. The narrow resonances at 561, 876~\&~1110~keV are shown with the vertical dotted lines. The astrophysical $S$--factor from NACRE--II~\cite{xu2013nacre} is also shown.}
\label{fig:12}
\end{figure}

%Comparison of NACRE and our S-factor
In addition to the thermonuclear reaction rate, we also calculated the astrophysical \textit{S}--factor. In Figure~\ref{fig:12} we present the astrophysical $S$--factor for \beag~with the individual resonant contributions. Our results agree well with the NACRE--II data, with the exception of the 2101~keV resonance which seems to be mis--placed to lower energies~\citep[see Figure 42 in Ref.][]{xu2013nacre}.

\section{Discussion and Conclusions}
\label{sec:6}

In the present work we performed the first inverse kinematics study of
the \beag~reaction to measure unknown resonance strengths at energies relevant to
$\nu p$--process nucleosynthesis. We report the first measurement of the 1155~ \&~ 1110~keV resonances with strengths of $1.73 \pm 0.25(stat.) \pm 0.40(syst.)$~eV  and $125 ^{+27}_{-25}(stat.) \pm 15(syst.)$~meV,
respectively. We also re--measured the 876~keV resonance strength ($\omega\gamma_{876}=3.00^{+0.81}_{-0.72} (stat.) \pm 0.61(syst.)$~eV) and
our result agrees with the measurement of~\citet{hardie1984resonant} ($\omega\gamma= 3.80(57)$~eV).

As we have also demonstrated in~\citet{psaltis6li}, the present
work shows that DRAGON is capable of handling
measurements in which the maximum recoil cone angle exceeds its acceptance,
after a systematic study of the BGO array efficiency and its transmission
using extensive {\scshape Geant} simulations. That opens a new
avenue for future experiments using DRAGON, that were previously thought to be
inaccessible due to large maximum recoil angles (\emph{e.g.}
$\isotope[18][]{O}(\alpha,\gamma)\isotope[22][]{Ne}$,
$\isotope[20][]{Ne}(\alpha,\gamma)\isotope[24][]{Mg}$ and others).

The new \beag~reaction rate is constrained to 9.4-10.7 \% for
T = 1.5-3~GK
which is sufficient for nucleosynthesis calculations. The effect of the rate,
along with other measured reactions relevant to nucleosynthesis in
neutrino--driven winds will be explored in a subsequent study. According
to the work of~\citet{wanajo2011uncertainties}, the \beag~reaction rate in the relevant energies can affect the number of the neutron--to--seed ratio $\Delta_n$, regulating the $\nu p$--process efficiency in synthesizing neutron--deficient isotopes. This is a particularly interesting result, since most recent self--consistent 3D core--collapse supernova simulations favour proton--rich conditions~\cite{bollig2021self}. In addition, a rigorous study of the astrophysical conditions of the proton--rich neutrino driven ejecta, and how they produce different nucleosynthesis outputs, using all the up--to--date nuclear physics input is desired.

The intense $\isotope[7][]{Be}$ RIBs produced with carbide targets can be
utilized for more demanding experiments, such as $\isotope[7][]{Be}(p,\gamma)$
and $\isotope[7][]{Be}$ $\alpha$--scattering. Pure graphite targets bombarded
by protons at $100~\mu A$ (or a UC$_x$ target at $40~\mu A$) could produce as
much as $10^{10}$~s$^{-1}$ of $\isotope[7][]{Be}$.\newline

\section*{Acknowledgements}

The authors gratefully acknowledge the beam delivery and
ISAC operations groups at TRIUMF. In particular, we thank
F. Ames, T. Angus, A. Gottberg, S. Kiy, J. Lassen
and O. Shelbaya for all their help
during the experiment. The authors thank the anonymous referee for useful comments that improved the manuscript. The core operations of TRIUMF
are supported via a contribution from the federal government through the
National Research Council of Canada, and the Government of British Columbia
provides building capital funds. Authors from McMaster University
are funded by the National Sciences
and Engineering Research Council of Canada (NSERC). DRAGON is funded through
NSERC grant SAPPJ-2019-00039. AP also acknowledges support from the Deutsche Forschungsgemeinschaft (DFG, German Research Foundation)-Project No. 279384907-SFB 1245, and the State of Hesse within the Research Cluster ELEMENTS (Project ID 500/10.006).
 RL and CM acknowledge support from the U.S. Department of Energy, Office of Science,
Office of Nuclear Physics under Grant Nos. DE-SC0017799 and DE-FG02-97ER41042.
CRB, RG and SP acknowledge support from the U.S. Department of Energy, under grants number DE-FG02-88ER40387 and DE-NA0003883.
Authors from
Colorado School of Mines acknowledge
support from U.S. Department of Energy Office of Science DE-FG02-93ER40789
grant. Authors from the UK are supported by the Science and Technology
Facilities Council (STFC).
This work benefited from discussions at the ``Nuclear Astrophysics at Rings and Recoil Separators" Workshop
supported by the National Science Foundation under Grant No.
PHY-1430152 (JINA Center for the Evolution of the Elements).

%\nocite{*}
\bibliographystyle{apsrev4-2}
\bibliography{apssamp}% Produces the bibliography via BibTeX.

\end{document}